\documentclass{article}

\usepackage[utf8]{inputenc}

\usepackage[T1]{fontenc}
\usepackage{parskip}
\usepackage{ragged2e}
\usepackage{enumitem}
\usepackage{float}
\usepackage{lscape}
\usepackage[colorinlistoftodos]{todonotes}
\usepackage{rotating}

\usepackage{fancyhdr}
\usepackage{etoolbox}
\usepackage{geometry}
 \newcommand{\tmpx}{}

\fancypagestyle{sec}{\lhead{\tmpx}}

\usepackage{amsmath}
\usepackage{amsfonts}
\usepackage{amssymb}
\usepackage{mathtools}
\usepackage{amsthm}
\usepackage{bbm}
\usepackage{nccmath} 
\usepackage{scalerel}
\usepackage{actuarialsymbol}
\usepackage{mleftright}
\usepackage{mathrsfs}

\usepackage{multirow}
\usepackage{multicol}
\usepackage{tabularx}

\usepackage{graphicx}
\usepackage[table]{xcolor}
\usepackage{array}
\usepackage{subcaption}

\usepackage{xurl}
\usepackage{thmtools, thm-restate}
\usepackage{enumitem}
\usepackage{titlesec}
\usepackage[absolute,overlay]{textpos}

\usepackage[sectionbib, round]{natbib}
\usepackage[colorlinks, linkcolor = black,  linkcolor = blue,
urlcolor  = blue,
citecolor = blue,
anchorcolor = blue]{hyperref}
\usepackage{etoolbox}
\usepackage{bookmark}

\usepackage{graphicx}
\usepackage{tikz}
\usetikzlibrary{arrows.meta,positioning,calc}

\newcommand{\rmd}{\mathrm{d}}

\definecolor{MyGreen}{RGB}{0,153,0}

\titleformat{\subsubsection}
{\normalfont\normalsize\bfseries}{\thesubsubsection}{1em}{}
\titlespacing*{\subsubsection}
{0pt}{3.25ex plus 1ex minus .2ex}{1.5ex plus .2ex}
 
 \titleformat{\paragraph}[runin]
{\itshape}{\theparagraph .}{1em}{}

\newcommand{\stripthanks}[1]{%
  \begingroup
  \let\thanks\@gobble
  #1%
  \endgroup
}

\allowdisplaybreaks

\makeatletter
\newcounter{author}

\newcommand{\authorclean}[1]{%
  \csgdef{author@clean@\the\c@author}{#1}%
}

\renewcommand*\author[1]{%
  \stepcounter{author}%
  \csgdef{author@full@\the\c@author}{#1}%
  \ifcsundef{author@clean@\the\c@author}{%
    \def\temp{#1}%
    \edef\temp{\unexpanded\expandafter{\temp}}%
    \csgdef{author@clean@\the\c@author}{\temp}%
  }{}%
  \ifnum\c@author=1
    \gdef\@author{#1}%
  \else
    \xdef\@author{\unexpanded\expandafter{\@author\and#1}}%
  \fi
}

\newcommand*\email[1]{%
  \csgdef{email@\the\c@author}{#1}}

\newcommand*\orcid[1]{%
  \csgdef{orcid@\the\c@author}{#1}}
  
\newcommand*\address[1]{%
  \csgdef{address@\the\c@author}{#1}}

\AtEndDocument{%
  \xdef\author@count{\the\c@author}%
  \c@author=1
  \print@authors}

\newcommand*\print@authors{%
  \ifnum\c@author>\author@count
  \else
    \print@author{\the\c@author}%
    \advance\c@author by 1
    \expandafter\print@authors
  \fi}

\newcommand*\print@author[1]{%
  \par\medskip
  \begin{tabular}{@{}l@{}}%
    \textsc{\csuse{author@clean@#1}}\\
    \csuse{address@#1}\\
    \textit{E-Mail}: \href{mailto:\csuse{email@#1}}{\csuse{email@#1}}\\
    \textit{ORCiD}:
    \href{\csuse{orcid@#1}}{\csuse{orcid@#1}}
  \end{tabular}}

\makeatother

\numberwithin{equation}{section}

\theoremstyle{theorem}
\newtheorem{definition}{Definition}
\numberwithin{definition}{section}
\theoremstyle{theorem}
\newtheorem{theorem}{Theorem}
\numberwithin{theorem}{section}
\theoremstyle{plain}
\newtheorem{proposition}{Proposition}
\numberwithin{proposition}{section}
\theoremstyle{plain}

\numberwithin{corollary}{section}
\newtheorem{lemma}{Lemma}
\numberwithin{lemma}{section}
\theoremstyle{plain}
\newtheorem{remark}{Remark}
\numberwithin{remark}{section}
\mathtoolsset{showonlyrefs}
\usepackage{etoolbox,xparse}

\numberwithin{example}{section}

\providecommand{\keywords}[1]
{
  \small	
  \textbf{\textit{Keywords---}} #1
}

\author{Kira Henshaw$^1$}
\authorclean{Kira Henshaw}
\address{\textit{Institute for Financial and Actuarial Mathematics} \\ \textit{Department of Mathematical Sciences} \\
\textit{University of Liverpool} \\ \textit{Liverpool, United Kingdom}}
\email{khenshaw@liverpool.ac.uk}
\orcid{https://orcid.org/0000-0002-9018-7993}
\author{Jorge Ramirez$^2$
}
\authorclean{Jorge Ramirez}
\address{\textit{Departamento de Matem\'aticas} \\ \textit{Universidad Nacional de Colombia} \\ \textit{Sede Medell\'in} \\ \textit{Medell\'in, Colombia}}
\email{jmramirezo@unal.edu.co}
\orcid{https://orcid.org/0000-0002-0402-7484}
\author{Jos\'e Miguel Flores-Contró$^3$}
\authorclean{Jos\'e Miguel Flores-Contró}
\address{\textit{Institute of Statistics, Biostatistics and Actuarial Science – ISBA} \\ \textit{Louvain Institute of Data Analysis and Modeling – LIDAM} \\
\textit{Catholic University of Louvain} \\ \textit{Louvain-la-Neuve, Belgium}}
\email{jose.flores@uclouvain.be}
\orcid{https://orcid.org/0000-0001-9332-8811}
\author{Sooie-Hoe Loke$^4$}
\authorclean{Sooie-Hoe Loke}
\address{\textit{Department of Mathematical Sciences} \\ \textit{Middle Tennessee State University} \\
\textit{Murfreesboro, Tennessee} \\ \textit{United States of America}}
\email{sooiehoe.loke@mtsu.edu}
\orcid{https://orcid.org/0000-0001-9021-1562}
\author{Enrique A. Thomann$^5$}
\authorclean{Enrique A. Thomann}
\address{\textit{Department of Mathematics} \\
\textit{Oregon State University} \\ \textit{Corvallis, Oregon} \\ \textit{United States of America}}
\email{thomann@math.oregonstate.edu}
\orcid{https://orcid.org/0000-0003-0972-1659}
\author{Corina Constantinescu$^1$}
\authorclean{Corina Constantinescu}
\address{\textit{Institute for Financial and Actuarial Mathematics} \\ \textit{Department of Mathematical Sciences} \\ \textit{University of Liverpool} \\ \textit{Liverpool, United Kingdom}}
\email{c.constantinescu@liverpool.ac.uk}
\orcid{https://orcid.org/0000-0002-5219-3022}

\title{On the Impact of Insurance on Households Susceptible to Random Proportional Losses: An Analysis of Poverty Trapping}

\begin{document}


\date{\vspace{-2ex} \footnotesize{
    $^1$Institute for Financial and Actuarial Mathematics, University of Liverpool, Liverpool, United Kingdom\\%
    $^2$Departamento de Matem\'aticas, Universidad Nacional de Colombia, Sede Medell\'in, Medell\'in, Colombia\\%
    $^3$Institute of Statistics, Biostatistics and Actuarial Science (ISBA), Louvain Institute of Data Analysis and Modeling (LIDAM), Catholic University of Louvain, Louvain-la-Neuve, Belgium\\%
   $^4$Department of Mathematical Sciences, Middle Tennessee State University, Murfreesboro, Tennessee, United States of America\\%
    $^5$Department of Mathematics, Oregon State University, Corvallis, Oregon, United States of America\\%
}}

\maketitle

\vspace{-3ex}

\begin{abstract}
The trapping probability is studied for households\rq \ capital assuming that losses are proportional to the accumulated capital. We consider households without insurance and those purchasing proportional insurance. Parameter conditions under which poverty trapping is not certain in both cases are derived. For the uninsured model, a new closed-form expression for the trapping probability is obtained when the remaining proportion of capital after a loss has a beta distribution, showing that the trapping probability decays algebraically as the initial capital increases. When insurance is purchased and the remaining proportion is uniformly distributed, the trapping probability is computed by solving a non-local integro-differential equation. Our results show that insurance can prevent otherwise certain trapping in some cases, although it may increase trapping risk for households with initial capital barely above the poverty line, while reducing trapping risk for households with higher initial capital.
\small{{{JEL classification}}: G220; G520; O120.}\\

\keywords{Microinsurance; poverty traps; trapping probability; risk processes; proportional losses; proportional microinsurance.}

\end{abstract}

\section{Introduction} \label{Introduction-Section1}
\subsection{Background}
Vulnerable non-poor households, i.e., those living just above the poverty line, are highly susceptible to falling into poverty, particularly following a financial loss. Understanding this vulnerability and the nature of their loss experience is essential for designing effective poverty-reduction policies and advancing Sustainable Development Goal 1 (SDG 1): No Poverty \citep{Book:UnitedNations2015}. An important indicator of financial stability is capital. In the vulnerable non-poor household setting, where monetary wealth is often limited, capital should be understood broadly to include the resources that enable production, whether for trade or self-sustaining purposes. This may include land, property, physical and human capital, with health a form of capital in extreme cases where sufficient health services and food accessibility are not guaranteed \citep{Article:Dasgupta1997}. The threat of catastrophic loss events is of great concern, particularly under this broad definition of capital. For example, vulnerable non-poor households are predominantly engaged in agricultural work and are exposed, among many other things, to natural disasters in the form of floods and droughts. In contrast to losses relating to health, life or death, agricultural losses can immediately eliminate a high proportion of a household\rq s ability to produce through loss of land and livestock, irrespective of their level of capital.

In this article, we study the behaviour of households\rq \ capital under the assumption of proportional capital loss experience. Proportionality in loss experience captures the exposure of households of all capital levels to both catastrophic and low severity loss events. This is particularly relevant in the vulnerable non-poor household setting, where, in addition to infrequent but serious events such as natural disasters, more common events such as hospital admissions and household deaths, can be detrimental. To do this, we adopt the ruin-theoretic approach proposed in \cite{Article:Kovacevic2011}, by using a risk process with deterministic growth and multiplicative losses to model the capital of a household. At capital loss events, accumulated capital is reduced by a random proportion of itself, rather than by an amount of random value, as in \cite{Article:Flores-Contro2024}. Processes of this structure are typically referred to as growth-fragmentation or growth-collapse processes, characterised by their growth in between the random collapse times at which downwards jumps occur. In these models, the randomly occurring jumps have a random size which is dependent on the state of the process immediately before the jump.

Our aim is to derive the probability that a household ever falls below the poverty line, where this probability mimics an insurer\rq s infinite-time ruin probability. To the best of our knowledge, \cite{Article:Kovacevic2011} and \cite{Article:Flores-Contro2024} are the only previous studies to examine this problem from a ruin-theoretic perspective. As in these earlier works, in this article, we consider the probability under two frameworks, one in which the household has no insurance coverage, and the other in which they are proportionally insured. We introduce insurance to assess its effectiveness as a measure of poverty reduction. Aligning with the low-income environment, proportional coverage is assumed to be provided through an \textit{inclusive insurance} product, specifically designed to cater for those excluded from traditional insurance services or without access to alternative effective risk management strategies. This type of product, targeted towards low-income populations, is commonly referred to as \textit{microinsurance} \citep{Book:Churchill2006, Book:Churchill2012}. In \cite{Article:Flores-Contro2024}, the risk process with deterministic growth and random-valued losses is instead used to assess the impact of government premium subsidy schemes on the probability of falling below the poverty line.

\subsection{Literature Review}
Although important, we do not consider the behaviour of a household\rq s capital below the poverty line. Households that live or fall below the poverty line are said to be in a poverty trap, where a poverty trap is a state of poverty from which it is difficult to escape without external help. Poverty trapping is a well-studied topic in development economics (the interested reader may refer to \cite{Inbook:Aghion2005}, \cite{Book:Bowles2006}, \cite{Article:Kraay2014}, \cite{Article:Barrett2016} and references therein for further discussion; see \cite{Inbook:Matsuyama2008} for a detailed description of the mechanics of poverty traps), however, for the purpose of this study, we use the term \lq \lq trapping\rq \rq \ only to describe the event that a household falls into poverty, focusing our interest on vulnerable non-poor households with initial capital levels slightly above the poverty line.

In \cite{Article:Kovacevic2011}, estimates of the infinite-time trapping 
probability (the probability that a household eventually falls into the 
poverty trap)
are obtained numerically. The authors also argue, by 
drawing an analogy with classical ruin theory, for the existence of a 
Lundberg coefficient and the resulting exponential decay of the trapping 
probability as the initial capital increases. 
\cite{Article:Azais2015} perform further numerical analysis of the same 
model, discussing applications to the capital setting of 
\cite{Article:Kovacevic2011} and to population dynamics, where the poverty 
line denotes extinction. In neither study is an analytical expression for 
the trapping probability obtained.

Our analysis leads to a different conclusion regarding the asymptotic 
behaviour of the model in \cite{Article:Kovacevic2011}. When the remaining 
proportion of capital after each loss event follows a 
$\mathrm{Beta}(\alpha,1)$ distribution, we derive a 
closed-form expression for the infinite-time trapping probability and show 
that it decays algebraically, rather than exponentially, as the initial 
capital tends to infinity. Thus, the Lundberg-type exponential decay 
proposed in \cite{Article:Kovacevic2011}, which relies on the existence of a positive solution to the corresponding Lundberg equation, does not hold for this model under the Beta$(\alpha,1)$ specification. 
This distinction is important for the proportional-loss model: unlike 
under additive losses, increasing initial capital does not reduce the 
trapping probability at an exponential rate.

\color{black}
A related issue concerns the conditions under which the infinite-time 
trapping probability is non-trivial. As in classical ruin theory, a 
meaningful analysis requires identifying parameter regimes in which 
trapping is not certain (see Remark 1.1 in \cite{Book:Mandjes2023}). 
Using an embedded process at successive loss times, we derive a 
corresponding \textit{net profit condition} for both the uninsured and insured 
capital models. The condition separates the parameter regimes in which trapping occurs with probability one from those in which the trapping probability vanishes as initial capital increases.

The insured model presents a different analytical difficulty. When 
proportional insurance coverage is purchased by the household, the 
trapping probability satisfies a non-local integro-differential equation. 
For uniformly distributed remaining proportions of capital, we transform 
this equation into a delay differential equation and develop a recursive 
layer  construction based on Sturm--Liouville methods. The constructive nature of the method allows for an analytic representation of the trapping probability that can be evaluated numerically.

The proportional nature of the capital losses distinguishes the resulting process from classical insurance surplus models with additive claims (see, e.g., \cite{Book:Asmussen2010} and \cite{Book:Mandjes2023}). Although the proportional loss factors are independent, the corresponding absolute losses depend on the capital level immediately before each loss event and therefore inherit the state dependence of the process. In addition, only the portion of household capital above the poverty line grows exponentially. These features lead to infinitesimal generators for the risk process that differ substantially from those arising in classical ruin models.

A separate literature that examines insurance and poverty trapping from a non-ruin perspective has been undertaken through application of multi-equilibrium models and dynamic stochastic programming \citep{Article:Chantarat2017, Inbook:Barrett2016, Article:Carter2018,Article:Liao2020,Article:Janzen2020,Inbook:Kovacevic2021}. With the exception of the latter, each of these studies considers the impact of subsidisation and the associated cost to the subsidy provider. \cite{Article:Will2021} and \cite{Article:Henshaw2023} extend the problem to the group-setting, assessing the impact of risk-sharing on the trapping probability. \cite{Article:Will2021} undertake a simulation-based study and \cite{Article:Henshaw2023} propose a Markov modulated stochastic dissemination model of group wealth interactions, using a bivariate normal approximation to calculate the trapping probability.

Notably, \cite{Article:Kovacevic2011}, \cite{Article:Liao2020} and \cite{Article:Flores-Contro2024} suggest that purchase of insurance and the associated need for premium payment increases the risk of falling below the poverty line for the most vulnerable. Barriers to microinsurance penetration that exist due to constraints on product affordability resulting from fundamental features of the microinsurance environment likely contribute to such observations; see \cite{Article:Eling2014} for a detailed overview on the determinants of microinsurance demand. Limited consumer financial literacy and experience, product accessibility and data availability, are examples of the unique characteristics that must be accounted for when designing effective and affordable microinsurance products. Through our analysis, we further investigate the case of proportional loss experience to assess the associated implications on the affordability of insurance.

\cite{Article:Janzen2020} optimise the level of insurance coverage across the population, observing that those in the neighbourhood of the poverty line do not optimally purchase insurance (without subsidies), instead suppressing their consumption and mitigating the probability of falling into poverty. This aligns with the increase in the trapping probability observed in the aforementioned studies, when those closest to the poverty line purchase insurance. Similarly, \cite{Inbook:Kovacevic2021} derive the retention rate process that maximises the expected discounted capital, by allowing adjustments in the retention rate of the policyholder after each capital loss throughout the lifetime of the insurance contract. In this article, however, the proportion of insurance coverage and the choice to insure is fixed across the population, as in \cite{Article:Kovacevic2011}, \cite{Article:Chantarat2017} and \cite{Article:Flores-Contro2024}. Our comparison is therefore conditional on a given insurance contract; we do not model optimal insurance demand or household welfare.

\subsection{Contribution}

Our contribution is fourfold. First, we derive a random recurrence 
relation for an auxiliary capital process associated with the original 
formulation of \cite{Article:Kovacevic2011}. This representation provides 
an alternative characterisation of the trapping event and leads naturally 
to a net profit condition (NPC) analogous to that of classical ruin theory. The 
condition separates parameter regimes in which trapping occurs with 
probability one from those in which the trapping probability converges to 
zero as initial capital increases, without imposing a distributional form for the remaining proportion of capital.

Second, we specialize to the uninsured model with $\mathrm{Beta}(\alpha,1)$-distributed remaining proportions of capital and derive a closed-form expression for the trapping probability. Unlike the preceding probabilistic characterization and NPC, this distributional assumption is introduced to obtain an explicit analytical solution, which then allows us to determine the asymptotic behaviour of the trapping probability.

Third, whereas \cite{Article:Kovacevic2011} argue for exponential decay through the 
existence of a Lundberg coefficient, we show that the trapping probability 
instead decays algebraically, provided the NPC is satisfied. Thus, the asymptotic behaviour of the 
proportional-loss model differs fundamentally from that of classical 
additive-loss risk processes.

Finally, for the proportionally insured model with uniformly distributed 
remaining proportions of capital, we transform the non-local 
integro-differential equation into a delay differential equation and 
develop a layer method that constructs its solution recursively over 
successive intervals. The resulting solution also allows us to compare the insured and uninsured households across initial capital levels. We find that proportional insurance can increase trapping risk close to the poverty threshold, but reduce it at higher capital levels and, in some parameter regimes, change the process from certain eventual trapping to a non-trivial trapping probability.

The remainder of this paper is structured as follows. Section \ref{TheCapitalModel-Section2} introduces the capital growth model under both the uninsured and insured settings. Section \ref{TheTrappingProbabilityandtheNetProfitCondition-Section3} develops an alternative characterisation of the trapping event and derives the corresponding net profit conditions. Section \ref{AnalyticalApproachtoTrappingProbabilitiesviaInfinitesimalGenerators-Section4} derives the trapping probabilities, including the closed-form solution for the uninsured model and the recursive layer method for the insured model. Section \ref{Discussion-Section5} compares additive and multiplicative losses and examines the impact of proportional insurance. Concluding remarks are provided in Section \ref{Conclusion-Section6}. Most proofs of the results in this paper
are provided in Appendix \ref{AppendixB:MathematicalProofs}.

\section{The Capital Model} \label{TheCapitalModel-Section2}
This section presents the stochastic capital models considered throughout the paper. For completeness, we describe both the uninsured and insured formulations that form the basis of our subsequent analysis. The notation and assumptions established here will be used throughout the remainder of the article.

\subsection{Uninsured Model}
\label{TheCapitalModel-Section2-Subsection1}

We begin by recalling the uninsured capital model introduced by \cite{Article:Kovacevic2011}.

Consider a household with accumulated capital $\{X_{0, t}\}_{t\geq 0}$, with the subindex  \lq \lq $0$\rq\rq \ connoting here the absence of insurance. Under the basic assumption that the household has no loss experience, its growth in accumulated capital is given by 

\begin{align}
    \frac{dX_{0,t}}{dt}=r_0 \cdot\left[X_{0,t}-x_{0}^{*}\right]^{+},
    \label{TheCapitalModel-Section2-Equation1}
\end{align}

where $r_{0}>0$ is the capital growth rate and $x_{0}^{*}>0$ is the threshold below which a household lives in poverty. Here, we denote $[x]^{+}=\max(x,0)$. 

The dynamics in \eqref{TheCapitalModel-Section2-Equation1} are built on the assumption that a household\rq s income $I_{t}$ is split into consumption $C_{t}$ and savings $S_{t}$ (or investments), such that at time $t$,

\begin{align}
    I_t=C_t+S_t,
    \label{TheCapitalModel-Section2-Equation2}
\end{align}

where consumption is an increasing function of income:

\begin{align}
    C_{t}= \begin{cases} I_{t}  \hspace{2.65cm} \textit{ if } \hspace{0.25cm} I_{t}\leq I^{*},\\ I^{*}+a\left(I_{t}-I^{*}\right) \hspace{0.5cm} \textit{ if } \hspace{0.25cm} I_{t}> I^{*}, \end{cases}
    \label{TheCapitalModel-Section2-Equation3}
\end{align}

for some household rate of consumption $0<a<1$. The critical point below which a household consumes all of its income, with no facility for savings or investment, is denoted by $I^*$. The accumulated capital is assumed to grow proportionally to the level of savings, such that

\begin{align}
    \frac{dX_{0,t}}{dt}=cS_{t},
    \label{TheCapitalModel-Section2-Equation4}
\end{align}

where $0<c<1$ is the household\rq s rate of savings. Income is generated through the accumulated capital, such that
$    I_{t}=bX_{0,t},
$
with $b>0$ denoting the household\rq s income generation rate. Combining \eqref{TheCapitalModel-Section2-Equation2}, \eqref{TheCapitalModel-Section2-Equation3} and \eqref{TheCapitalModel-Section2-Equation4} gives exactly the dynamics in \eqref{TheCapitalModel-Section2-Equation1} with

\begin{align}
    r_0 = (1-a) \cdot b \cdot c \quad \text{ and } \quad x_0^*=\frac{I^*}{b}.
    \label{TheCapitalModel-Section2-Equation6}
\end{align}

The notion of a household in this model setting may be extended for consideration of poverty trapping within economic units such as community groups, villages and tribes, in addition to the traditional household structure. Reflecting the ability of a household to produce, the level of accumulated capital of a household up to time $t$, $X_{0,t}$, is composed of land, property, physical and human capital. The  poverty threshold $x_{0}^{*}$ represents the amount of capital required to forever attain a critical level of income below which a household would not be able to sustain their basic needs, facing elementary problems relating to health and food security. We refer to this threshold as the \textit{critical capital} or the \emph{poverty line}. Since \eqref{TheCapitalModel-Section2-Equation1} is positive for all levels of capital greater than the critical capital, all points less than or equal to $x_{0}^{*}$ are stationary, the level of capital remains constant if the critical capital is not met. In this basic model, stationary points below the critical capital are not attractors of the system if the initial capital exceeds $x_{0}^{*}$, in which case the capital process grows exponentially with rate $r_{0}$.

In line with \cite{Article:Kovacevic2011}, we expand the dynamics of \eqref{TheCapitalModel-Section2-Equation1} under the assumption that households are susceptible to the occurrence of capital losses such as those highlighted in Section \ref{Introduction-Section1}, including severe illness, the death of a household member or breadwinner and catastrophic events such as droughts, floods and earthquakes. The occurrence of loss events is assumed to follow a Poisson process with intensity $\lambda$, where the capital process follows the dynamics of \eqref{TheCapitalModel-Section2-Equation1} in between events. 

On the occurrence of the $i$th loss, the capital process experiences an instantaneous  downwards jump from $X_{0,T_i}$ to  $X_{0, T_i}\cdot Z_i$, where $Z_i\in[0,1]$ is the random proportion determining the remaining capital after loss $i$ and $X_{0, T_i}$ is the level of capital accumulated up to the time of loss $T_{i}$. The sequence $\{Z_{i}\}_{i=1}^{\infty}$  is a sequence of independent and identically distributed random variables with common distribution function $G_{Z}$, independent of the Poisson process. In contrast to \cite{Article:Flores-Contro2024} where random-valued losses are considered, the dynamics of the model do not allow for the possibility of negative capital due to the proportionality of loss experience.

The structure of the process in-between loss events is derived through solution of the first order ordinary differential equation (ODE) in \eqref{TheCapitalModel-Section2-Equation1}. The stochastic capital process with deterministic exponential growth and multiplicative losses is then formally defined as follows:

\begin{definition} \label{TheCapitalModel-Section2-Definition1}

Let $T_{i}$ be the $i$th event time of a Poisson process $\{N_{t}\}_{t\geq0}$ with parameter $\lambda$, where $T_{0}=0 .$ Let $0 \leq Z_{i} \leq 1$ be a sequence of i.i.d.\ random variables with distribution function $G_{Z}$, independent of the process $N_{t}$. For $T_{i-1} \leq t<T_{i}$, the stochastic growth process of the accumulated capital without insurance coverage $X_{0,t}$ is defined as

\begin{align}
    X_{0,t}=\begin{cases} \left(X_{0,{T_{i-1}}}-x_{0}^{*}\right) e^{r_{0} \left(t-T_{i-1}\right)}+x_{0}^{*} & \textit { if } \hspace{0.25cm} X_{0, T_{i-1}}>x_{0}^{*}, \\ X_{0,T_{i-1}} & \textit{otherwise}.
    \end{cases}
     \label{TheCapitalModel-Section2-Equation7}
\end{align}

At the jump times $t = T_{i}$, the capital process without insurance coverage is given by

\begin{align}
    X_{0, T_{i}}=\begin{cases} \left[\left(X_{0, T_{i-1}}-x_{0}^{*}\right) e^{r_{0} \left(T_{i}-T_{i-1}\right)}+x_{0}^{*}\right] \cdot Z_{i} & \textit { if } \hspace{0.25cm} X_{0, T_{i-1}}>x_{0}^{*}, \\ X_{0, T_{i-1}} \cdot Z_{i} & \textit{otherwise}.
    \end{cases}
     \label{TheCapitalModel-Section2-Equation8}
\end{align}

\end{definition}

Definition \ref{TheCapitalModel-Section2-Definition1} indicates that the capital level of the household follows a piecewise deterministic Markov process \citep{Article:Davis1984, Book:Davis1993} of compound Poisson-type. Namely, the capital evolves deterministically in-between the randomly occurring jump times at which capital losses occur.
Consequently, the infinitesimal generator of the capital process without insurance coverage is given by
\begin{align}
    \mathcal{A}_{0}[h](x)=r_0(x-x_0^*)h'(x)+\lambda\int_0^1[h(x\cdot z)-h(x)]\rmd G_{Z}(z), \quad x \geq x^*_{0},
    \label{InfinitesimalGeneratorNoInsurance}
\end{align} 

where $h$ is any function in the domain of the generator $\mathcal{A}_{0}$.

\subsection{Insured Model} \label{TheCapitalModel-Section2-Subsection2}

In line with \cite{Article:Kovacevic2011} and \cite{Article:Flores-Contro2024}, we now extend the model by assuming that households purchase insurance coverage. 

Let $\mathcal{R}:[0,1] \rightarrow[0,1]$ denote the retained loss function, which satisfies $0 \leq \mathcal{R}(l) \leq 1$ and $\mathcal{R}(0)=0$. This function specifies the portion of a loss borne by the household when a loss of size $l = 1-z \in[0,1]$ occurs per unit of capital. Accordingly, after a loss event, the household\rq s capital changes from $X_{T_{i}}$ to $(1-\mathcal{R}(l)) \cdot X_{T_{i}}$ (note that we now omit the subscript \lq \lq 0\rq \rq \ when referring to the accumulated capital with insurance). The insurer determines the premium rate $p$ that the household must pay for the insurance coverage using the \textit{expected value principle}. That is,

\begin{align}
    p = (1+\theta) \cdot \lambda \cdot \mathbb{E}[1-Z_{i}-\mathcal{R}(1-Z_{i})],
    \label{TheCapitalModel-Section2-Equation9}
\end{align}

where $\theta > 0$ is the safety loading per unit of capital. Under this insurance coverage, the capital growth rate $r$ of an insured household must account for the premium payments, which are deducted from the income generation rate. Naturally, we assume that $b>p$ and this yields

\begin{align}
    r=(1-a) \cdot\left(b-p\right) \cdot c=
    r_0\left(\frac{b-p}{b}\right)\leq r_0.
    \label{TheCapitalModel-Section2-Equation10}
\end{align}

Moreover, the critical capital (or poverty line) is now given by

\begin{align}
    x^{*}=\left(\frac{b}{b-p}\right) x_0^{*} \geq x_0^{*},
    \label{TheCapitalModel-Section2-Equation11}
\end{align}

since $x_0^{*}=I^{*} / b$ and $x^{*}=I^* /\left(b-p\right)$, where 
$I^*$ denotes the critical income.

Clearly, the capital model of an insured household has an analogous structure to that of Definition \ref{TheCapitalModel-Section2-Definition1}, with the remaining proportion of capital after each loss event instead denoted $Y_{i}$, where
\[Y_{i} := 1-\mathcal{R}(1-Z_{i}).\] 
As such, in between loss events, where $T_{i-1}\leq t<T_i$, the capital growth process follows \eqref{TheCapitalModel-Section2-Equation7}  with $r$ and $x^{*}$ replacing $r_{0}$ and $x_{0}^*$, respectively. At event times $t = T_{i}$, the process is given by 

\begin{align}
    X_{T_{i}}=\begin{cases} \left[\left(X_{T_{i-1}}-x^{*}\right) e^{r \left(T_{i}-T_{i-1}\right)}+x^{*}\right] \cdot Y_{i} & \textit { if } \hspace{0.25cm} X_{T_{i-1}}>x^{*}, \\ X_{T_{i-1}} \cdot Y_{i} & \textit{otherwise}.
    \end{cases}
     \label{TheCapitalModel-Section2-Equation12}
\end{align}

Motivated by the importance of designing simple, transparent, and easy-to-understand insurance products in the microinsurance context \citep{Book:Churchill2006, Book:Churchill2012, Article:Eling2014}, and by the fact that proportional insurance is generally considered to possess these characteristics \citep{Article:Lampaert2005}, we assume throughout this paper that the household acquires a quota-share proportional insurance coverage. Let $\kappa  \in[0,1] $ denote the proportion of the loss $l$ that is retained by the household so that the retained loss function takes the form 
\begin{align}
    \mathcal{R}(l)=\kappa \cdot l, 
    ~~~~~~{\textrm{so that}},
    ~~~~~~Y_{i}=1-\kappa(1-Z_{i}).
    \label{eq:defYi}
\end{align}

In this setting, the premium rate per unit of capital is given by
\begin{align}
    p=(1+\theta)\cdot \lambda \cdot(1-\kappa)  \cdot \left(1 -\mathbb{E}\left[Z_{i}\right]\right). 
    \label{TheCapitalModel-Section2-Equation14}
\end{align}
Note that $\kappa=0$ corresponds to the full insurance case, where the insurer bears all of the household\rq s losses. On the other hand, $\kappa=1$ represents the absence of insurance, in which case the capital model in \eqref{TheCapitalModel-Section2-Equation12}, together with the parameters $r$ and $x^{*}$, coincides exactly with the uninsured household model introduced earlier in Section \ref{TheCapitalModel-Section2-Subsection1}.

Thus, in the  case of proportional insurance considered in this paper, the infinitesimal generator for the insured capital process as defined in this section is given by
\begin{align}
    \mathcal{A}[h](x)=r\left(x-x^{*}\right) h^{\prime}(x)-\lambda h(x)+\lambda \int_{1-\kappa}^1 h(x \cdot y) \rmd G_Y(y),
\label{TheTrappingProbabilityandInfinitesimalGenerator-Section3-Equation18}
\end{align}
where
\[G_{Y}(y) = G_{Z}(1-(1-y) / \kappa)\] 
is the cumulative distribution function of $Y_{i}$, whose support is $[1-\kappa,1]$, and $h$ is any function which lies in the domain of the generator $\mathcal{A}$. 

\begin{remark} \label{TheCapitalModel-Section2-Remark1}

Up to this point, we have denoted the uninsured capital process and its associated parameters by $\{X_{0,t}\}_{t\geq0}$, $r_0$, and $x^*_0$  in order to distinguish them from their counterparts $\{X_{t}\}_{t\geq0}$, $r$, and $x^*$, arising in the insured capital model. To streamline the notation and improve expositional clarity, we henceforth suppress the subscript \lq\lq 0\rq \rq \ distinction and use a common notation for both models, except when an explicit comparison is required.
    
\end{remark}

\begin{figure}[]
	\begin{subfigure}[b]{0.5\linewidth}
 		\includegraphics[scale=0.6]{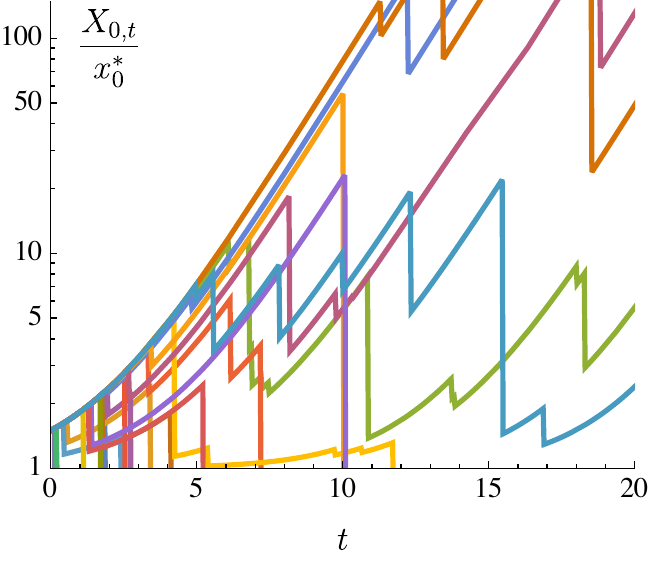}
		\caption{}
 		\label{TheCapitalModel-Section2-Figure1-a}
	\end{subfigure}
	\begin{subfigure}[b]{0.5\linewidth}
 		\includegraphics[scale=0.59]{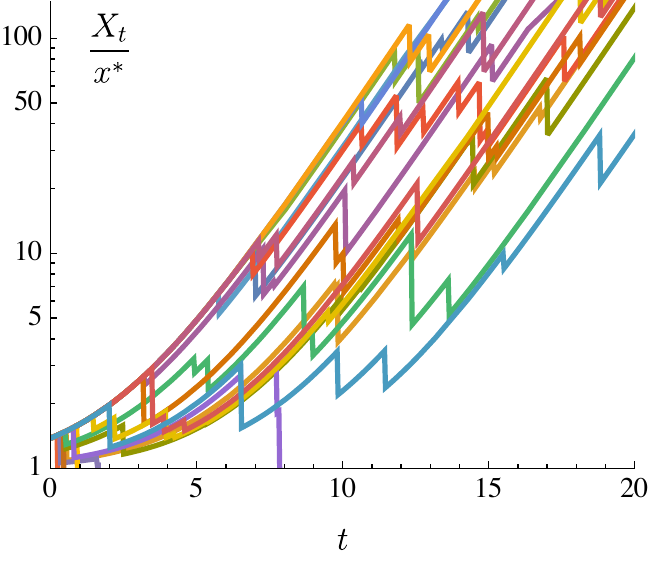}
		\caption{}
 		\label{TheCapitalModel-Section2-Figure1-b}
	\end{subfigure}
	\caption{Examples of paths of the capital process $X_{t}$ with initial capital $X_{0} = x = 1.5$,  $\lambda = 0.4$ and uniformly distributed $Z_i$. Note the logarithmic scale in the vertical axes. Trapping occurs for paths that cross the horizontal axes. Both plots have parameters $a = 0.1$, $b = 1.4$ and $c = 0.4$. Plot (a) shows the uninsured case with trapping probability $\psi(x) = 0.88$ as per Proposition \ref{AnalyticalApproachtoTrappingProbabilitiesviaInfinitesimalGenerators-Section4-Proposition1}. Plot (b) shows the insured case with $\theta = 0.1$ and $\kappa = 0.25$. The trapping probability can be estimated to be $\psi(x) \approx 0.35$ with the techniques of Section \ref{AnalyticalApproachtoTrappingProbabilitiesviaInfinitesimalGenerators-Section4-Subsection2}. }
     \label{TheCapitalModel-Section2-Figure1}
\end{figure}

Figure \ref{TheCapitalModel-Section2-Figure1} illustrates the capital dynamics both without and with insurance coverage. The parameters $r$ and $x^{*}$ are different for each setting, reflecting the presence or absence of insurance. In the insured case, the exponential growth rate $r$ is lower and losses are less severe; conversely, the uninsured case exhibits faster capital accumulation but more damaging shocks. Moreover, the poverty trap region is wider under insurance, as reflected by a larger critical capital $x^{*}$.

\section{Probabilistic Characterisation of the Capital Process}
\label{TheTrappingProbabilityandtheNetProfitCondition-Section3}

With the stochastic capital models now fully specified, we turn to the main objective of this paper, which is to study the probability that a household eventually falls below the poverty line under both uninsured and insured capital models. As in \cite{Article:Kovacevic2011} and \cite{Article:Flores-Contro2024},  we refer to the first time at which the capital falls below the poverty line as the \textit{trapping time}, defined by 

\begin{align}
    \tau:=\inf\left\{t\geq0:X_t<x^*\right\},
    \label{TheTrappingProbabilityandInfinitesimalGenerator-Section3-Equation1}
\end{align}

and define the corresponding  (infinite-time) \textit{trapping probability} as a function of the initial capital by  

\begin{align}
    \psi(x):= \mathbb{P}\left(\tau<\infty \mid X_0=x \right).\label{TheTrappingProbabilityandInfinitesimalGenerator-Section3-Equation2}
\end{align}
The infinite-time criterion is intended as a measure of long-run vulnerability rather than a short-term poverty forecast. 
Before studying the trapping probability analytically, it is useful to first understand the probabilistic structure of the capital process. In this section, we characterise the trapping time using an embedded discrete-time process and establish net profit conditions (analogous to that in classical ruin theory; see, e.g.,  \cite{Book:Asmussen2010} and \cite{Book:Mandjes2023}) for both the uninsured and insured models. These results identify when the infinite-time trapping probability is non-trivial (equal to one) and motivate the analytical treatment developed in Section \ref{AnalyticalApproachtoTrappingProbabilitiesviaInfinitesimalGenerators-Section4}.

\subsection{The Net Profit Condition in the Uninsured Model} \label{NetProfitConditionfortheUninsuredModel}

An alternative characterisation of the trapping time is developed in this section. Recall that, as noted in Remark \ref{TheCapitalModel-Section2-Remark1}, $x^*, r$, and $X_t$ denote the poverty line, capital growth rate, and capital model, respectively, in  both the uninsured or insured case. Likewise, in either case, we let $Y_i$ denote the remaining fraction of capital after the $i$th event.

First, observe that the trapping time (if finite) occurs at exactly one of the jump times $T_i$. Denote the capital process discretised at loss event times by 
\[
\tilde{X}_i:=X_{T_i}.
\]
Take $\tilde{X}_0=x$ and assume that $x>x^*>0$. Let $V_i:=T_i-T_{i-1}$ be the $i$th inter-arrival time, and rewrite \eqref{TheCapitalModel-Section2-Equation12}  as
$$\tilde{X}_i=\begin{cases} \left[\left(\tilde{X}_{i-1}-x^{*}\right) e^{rV_i}+x^{*}\right] \cdot Y_{i} & \textit { if } \hspace{0.25cm}\tilde{X}_{i-1}>x^{*}, \\ \tilde{X}_{i-1} \cdot Y_{i} & \textit{otherwise}.
    \end{cases}
   $$
Note that $V_i\sim \mathrm{Exp}(\lambda)$. The discretised process is sufficient to characterise the trapping time. However, the following auxiliary process admits a particularly convenient recursive representation. On the same probability space as $\{\tilde{X}_i\}_{i \in \mathbb{N}}$, define the auxiliary process $\{\widehat{X}_i\}_{i\in\mathbb{N}_{0}}$ by 
\begin{equation}\widehat{X}_i=\left[\left(\widehat{X}_{i-1}-x^{*}\right) e^{r V_i}+x^{*}\right]\cdot Y_i,
\label{eq:aux}
\end{equation}
with $\widehat{X}_0=x>x^*>0$. 
Note that the auxiliary process $\{\widehat{X}_i\}_{i\in\mathbb{N}_{0}}$
generalises the Markov chain introduced in Equation (18) of
\cite{Article:Kovacevic2011}, which is subsequently used there to argue
for the existence of a Lundberg coefficient. We take a different approach.
By exploiting the recursive structure of $\widehat{X}_i$ and applying the
strong law of large numbers, we obtain directly the net profit condition
that separates the regime of certain trapping from that in which the
trapping probability vanishes as the initial capital increases. 

\begin{remark} \label{interpauxHecapital}
    Equation  \eqref{eq:aux} admits a simple financial interpretation: the household\rq s capital at time $T_{i}$  consists of two components: first, a fraction $Y_{i}$ of the excess of capital over the critical capital level held at time $T_{i-1}$, accumulated over the period $[T_{i-1}, T_{i}]$ at a constant force of interest $r$; and a second, a deposit at time $T_{i}$ equal to a fraction $Y_{i}$ of the critical capital, i.e., $Y_{i}x^{*}$.
\end{remark}

\begin{figure}[H]
\centering
\includegraphics[scale=0.35 ]{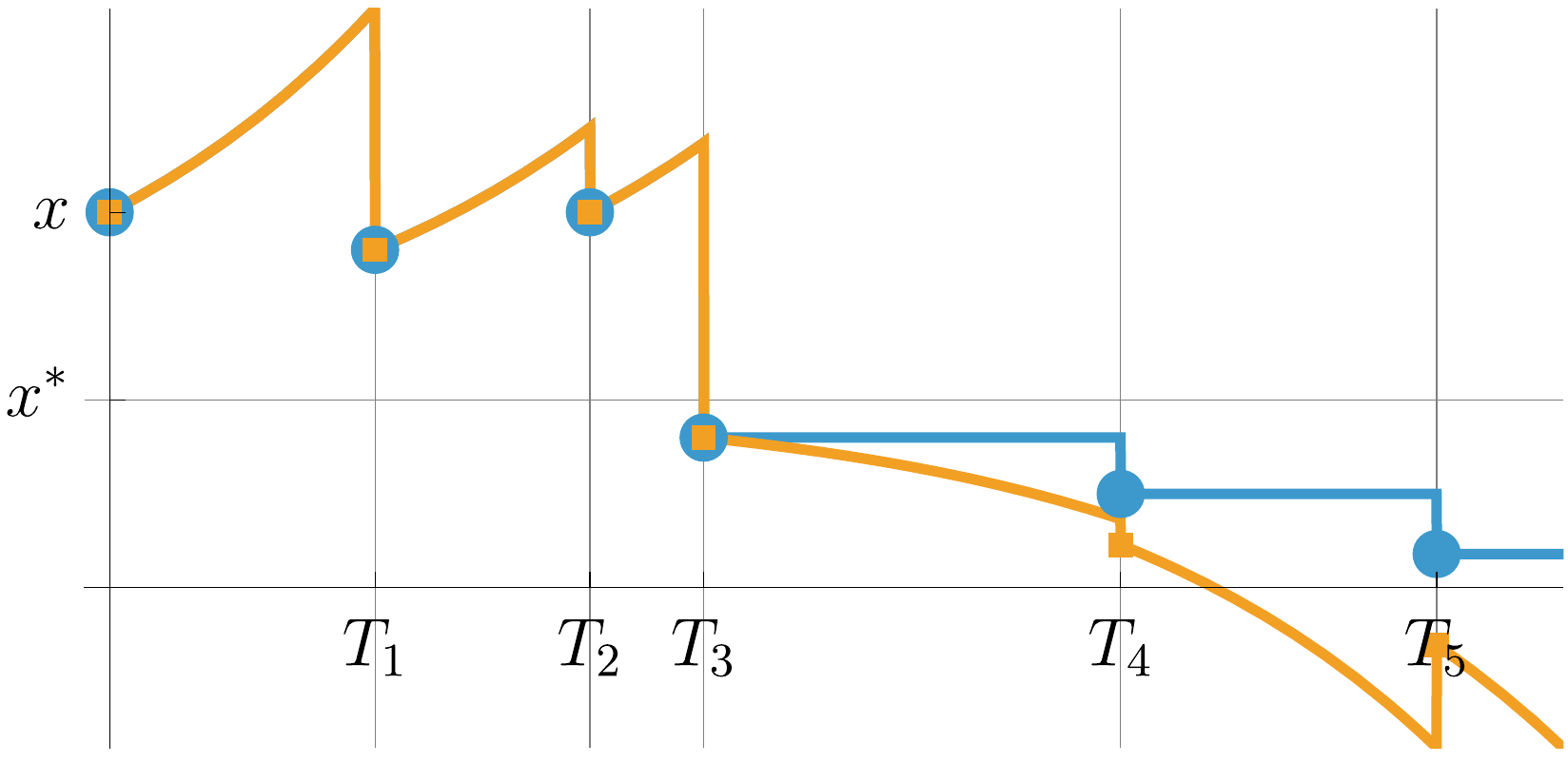}
\caption{Sample paths of the discretised process $\{\tilde{X}_{t}\}_{t\geq 0}$ labeled by blue dots and the auxiliary process $\{\widehat{X}_t\}_{t\geq 0}$ labeled by orange squares.}
\label{discVSaux}
\end{figure}

Figure \ref{discVSaux} shows examples of sample paths for both of these processes.  For every realisation, the two processes $\{\tilde{X}_i\}_{i\in\mathbb{N}_{0}}$ and $\{\widehat{X}_i\}_{i\in\mathbb{N}_{0}}$ are identical as long as they do not cross $x^*$. After crossing, the entire path of $\widehat{X}_i$ lies below the path of $\tilde{X}_i$ for all future time steps. Therefore, the two processes have the same trapping time and so we have established the following equivalencies for the trapping time $\tau$ in terms of the continuous capital process, discretised process, and auxiliary process: 
\begin{equation}\tau=\inf\left\{t\geq0:X_{t}<x^*\right\}=\inf\left\{T_i:\tilde{X}_i<x^*\right\}=\inf\left\{T_i:\widehat{X}_i<x^*\right\}.
\end{equation}

The trapping time can be characterised as follows:

\begin{proposition} 
\label{trappingtimecharacterization}
Let $\{\widehat{X}_i\}_{i\in\mathbb{N}_0}$ be defined by \eqref{eq:aux}.  For $k\ge 1$, let $\Pi_k=\prod_{j=1}^k  Y_j$ and set
$$M=\sum_{k=1}^\infty  \frac{1-Y_k}{\Pi_k} e^{-rT_k}.$$
 Then,
$$
\mathbb{P}(\tau<\infty \mid \widehat{X}_0 = x) = \mathbb{P}\left(\frac{x-x^*}{x^*} < M\right)
.$$
\end{proposition}

\begin{proof}
Let $R_i:=\widehat{X}_i-x^{*},$
 be the excess above the critical level at time $T_i$.  If $R_{i-1}>0$, we can rewrite \eqref{eq:aux} as
\begin{equation}
    R_i=Y_i e^{rV_i} R_{i-1}-(1-Y_i)x^*.
    \label{eqn:recursion}
\end{equation}
Iterating the recursion \eqref{eqn:recursion} yields
\[
R_i
=
e^{r T_i}\Pi_i
\left(
R_0
-
x^*\sum_{k=1}^i
\frac{1-Y_k}{\Pi_k} e^{-rT_k}
\right).
\]

Since \(\Pi_i>0\), the trapping time may be equivalently written as 
\begin{equation}
    \tau=\inf\{T_i:R_i< 0\}=\inf\left\{T_i: R_0< x^*\sum_{k=1}^i
\frac{1-Y_k}{\Pi_k}e^{-rT_k}\right\}.
\end{equation}
It follows that \begin{equation}
  \mathbb{P}(\tau<\infty \mid \widehat{X}_0 = x)=\mathbb{P}\left( \frac{R_0}{x^*}<\sup_{i\geq 1} \sum_{k=1}^i
\frac{1-Y_k}{\Pi_k} e^{-rT_k} \right).
\end{equation}
Since the summands are positive, 
\begin{equation}
    \sup_{i\geq 1} \sum_{k=1}^i
\frac{1-Y_k}{\Pi_k} e^{-rT_k} =\sum_{k=1}^\infty 
\frac{1-Y_k}{\Pi_k} e^{-rT_k}=M,
\end{equation}
completing the proof. 
\end{proof}

\begin{remark}
The quantity $M$ introduced in Proposition \ref{trappingtimecharacterization} has a natural interpretation: it represents the present value, when the force of interest is $r$, of a unit capital being available for investments at time $T_j, j=1, 2, \dots$, when the investment is subject to proportional losses at time $T_i, 1\le i\le j$ leaving fraction $Y_i$ of the capital available for investment.  Proposition \ref{trappingtimecharacterization} shows that trapping is avoided, if the initial excess capital $x-x^*$ is large enough to endow a fund that makes $x^*$ available at each time $T_j, j=1, 2, \dots.$
\end{remark}

Proposition \ref{trappingtimecharacterization} shows that the long-run behaviour of the capital process is governed by the multiplicative random factors $Y_j e^{rV_j}$. We therefore introduce the following notation. Let $A_j= Y_j e^{rV_j}, j=1,2,\dots$ be the i.i.d. random variables representing the remaining accumulation of a unit of capital under a force of interest $r$ over the inter-arrival time $V_i$. Then, 
$$
\frac{e^{-rT_k}}{\Pi_k} = \frac{e^{-rT_k}}{{\prod_{j=1}^k}Y_j} = \frac{1}{\prod_{j=1}^k A_j}.
$$

We are now in a position to state the following proposition:

\begin{proposition} \label{netprofitconditionproposition}
\begin{enumerate}
\item[(i)] If $\mathbb{E}[\ln \left(A_1\right)]>0$, then
$\psi(x)\to 0$ as $x \to \infty$. 
\item[(ii)] If $\mathbb{E}[\ln \left(A_1\right)]<0$,  then $\psi(x)=1$ for all $x>x^*$.
\end{enumerate}
\label{prop:npc}
\end{proposition}

\begin{proof}

By the strong law of large numbers,
$
\frac{1}{n} \sum_{j=1}^n \ln\left( A_j\right)
\to
\mathbb{E}[\ln \left(A_1\right)]$ almost surely. 
If $\mathbb{E}[\ln \left(A_1\right)]>0$, 
then \(\ln \left(\prod_{j=1}^n A_j\right)\to+\infty\), i.e. \(\prod_{j=1}^n A_j\) grows exponentially fast a.s. As such, $M$ is finite a.s. and it follows that
\[
\mathbb{P}(\tau<\infty \mid {X}_0 = x)
=
\mathbb{P}\left(
M\ge \frac{x-x^*}{x^*}
\right)\to 0,
\qquad\text{as }
x\to\infty.
\]
If instead $\mathbb{E}[\ln\left(A_1\right)]<0$, 
then $\ln \left(\prod_{j=1}^n A_j\right)\to-\infty$ and so $\prod_{j=1}^n A_j\to 0$ a.s.
The quantities \(1-Y_k\) are positive infinitely often a.s. which forces $M$ to diverge to \(+\infty\) a.s. Thus
$\mathbb{P}(\tau<\infty \mid {X}_0 = x)
=1$ in this case.
\end{proof}

Proposition \ref{prop:npc} identifies the balance between deterministic capital growth and stochastic proportional losses that governs the long-run behaviour of the capital process. When the average logarithmic accumulation is positive, sufficiently large initial capital makes trapping increasingly unlikely. Conversely, when the logarithmic accumulation is negative, every household eventually falls below the poverty threshold regardless of its initial capital.

In terms of the parameters of the capital process, we have $\mathbb{E}\left[\ln \left(A_1\right)\right]=\mathbb{E}\left[\ln \left(Y_1\right)\right]+ \frac{r}{\lambda}$ 
so that the condition for obtaining the non-trivial trapping probability becomes 
\begin{equation}
  \lambda   \mathbb{E}\left[\ln \left(Y_1\right)\right]+ r>0.
  \label{npc:uninsured}
\end{equation}

\begin{remark}
There is a close structural connection between the discretised model described above and the classical Cram\'{e}r-Lundberg model, originally introduced by \cite{Book:Lundberg1903} and \cite{Book:Cramer1930}. In the latter framework, an insurer receives premium income continuously while incurring claims of random size at stochastic arrival times. A central object of study in this setting is the infinite-time \textit{ruin probability}, defined as the probability that the insurer\rq s surplus ever becomes negative, leading to insolvency. This notion is strongly related to the trapping probability considered in the present work.
In classical ruin theory, the analysis of the infinite-time ruin probability begins with the formulation of the \textit{net profit condition}, which ensures that the insurer's expected premium income exceeds its expected claim outflows. Condition \eqref{npc:uninsured} plays an analogous role in our setting; accordingly, we adopt the same terminology and refer to it as the net profit condition. 
\end{remark}

The preceding result simplifies considerably when the remaining proportions of capital follow a Beta$(\alpha,1)$ distribution, a case that is analysed in greater detail later in the paper.

\begin{proposition}\label{TheTrappingProbabilityandInfinitesimalGenerator-Section3-Proposition2}

Consider the capital process of a household without insurance coverage as proposed in Definition \ref{TheCapitalModel-Section2-Definition1} with initial capital $x\ge x^{*}$, capital growth rate $r$, loss intensity $\lambda > 0$ and remaining proportions of capital $Z_i$ with distribution Beta$(\alpha,1)$. If the following net profit condition holds,
\begin{align}
    \rho := \frac{\lambda}{r} <\alpha,
    \label{npc:rho}
\end{align}
then $\lim_{x\to \infty} \psi(x) = 0$. 

\end{proposition}

\begin{proof}

With $Z_i \sim \textrm{Beta}(\alpha, 1)$, one has $\mathbb{E}[\ln(Z_i)]=\alpha\int_0^1\ln(z)z^{\alpha-1}\rmd z=-\tfrac{1}{\alpha}$.  Thus, \eqref{npc:rho} holds if and only if \eqref{npc:uninsured} is satisfied. Applying Proposition \ref{prop:npc} gives the desired result. 
\end{proof}

Since households face  trapping with probability one if the net profit condition is violated, our subsequent analysis in Section \ref{AnalyticalApproachtoTrappingProbabilitiesviaInfinitesimalGenerators-Section4}  focuses only on the region for which \eqref{npc:rho} holds. In particular, the assumptions made in Proposition \ref{TheTrappingProbabilityandInfinitesimalGenerator-Section3-Proposition2} allow for a fully explicit expression of the trapping probability as demonstrated in Section \ref{AnalyticalApproachtoTrappingProbabilitiesviaInfinitesimalGenerators-Section4-Subsection1}.  In the statement of this result, as in others presented throughout this article, the non-dimensional quantity $\rho=\tfrac{\lambda}{r}$ naturally appears. For clarity and consistency, we express our results in terms of $\rho$ whenever appropriate.

\subsection{The Net Profit Condition Under Proportional Insurance}
\label{TheTrappingProbabilityandInfinitesimalGenerator-Section3-Subsection3}

We now extend the analysis of Section \ref{NetProfitConditionfortheUninsuredModel} to households purchasing quota-share proportional insurance coverage as the one introduced in Section \ref{TheCapitalModel-Section2-Subsection2}. Although the framework remains unchanged, insurance modifies the effective capital accumulation through both premium payments and retained losses. Consequently, the corresponding net profit condition differs from that of the uninsured model, as given by the following proposition: 

\begin{proposition}

\label{TheTrappingProbabilityandInfinitesimalGenerator-Section3-Proposition3}
Consider the capital process of a household with proportional insurance coverage and proportionality factor $\kappa \in [0,1]$, defined by  
\eqref{TheCapitalModel-Section2-Equation12}. Furthermore, consider an initial capital $x\ge {x^{*}}$, capital growth rate $r$, loss intensity $\lambda > 0$ and remaining proportions of capital $Z_i$ with distribution Beta$(\alpha,1)$. The net profit condition in Proposition \ref{netprofitconditionproposition} is,

\begin{align}
    \frac{1}{\rho} = \frac{r}{\lambda} >  \frac{\kappa}{(\alpha+1)(1-\kappa)}{_2F_1}\left(1,\alpha+1;\alpha+2;-\frac{\kappa}{1-\kappa}\right),
    \label{TheTrappingProbabilityandInfinitesimalGenerator-Section3-Equation12}
\end{align}

where ${_2F_1}(\cdot)$ is the Gauss hypergeometric function  as defined in \eqref{AppendixA:DefinitionandSelectedPropertiesofGaussHypergeometricFunction-Subsection1-Equation1}. If \eqref{TheTrappingProbabilityandInfinitesimalGenerator-Section3-Equation12} holds, then $\lim_{x\to \infty} \psi(x) = 0$, otherwise trapping is certain.

\end{proposition}

\begin{proof}

By Proposition \ref{prop:npc}, the net profit condition for the capital process with proportional insurance coverage can be written as $    \mathbb{E}\left[\ln \left(Y_i e^{r V_i}\right)\right]>0$, where $Y_{i} = 1-\kappa(1-Z_{i})$ as per \eqref{eq:defYi}. By independence of $Y_i$ and $V_i$, the above inequality reduces to 
$    \frac{r}{\lambda}>-\mathbb{E}\left[\ln \left(Y_i\right)\right].
$ For $Z_i\sim \text{Beta}(\alpha,1)$, using integration by parts, we have
\begin{align}
    \mathbb{E}[\ln(1-\kappa(1-Z_i))]=-\kappa\int_0^1(1-\kappa+\kappa z)^{-1}z^{\alpha}\rmd z.
    \label{TheTrappingProbabilityandInfinitesimalGenerator-Section3-Equation16}
\end{align}
The above is an integral representation of a Gauss hypergeometric function \eqref{AppendixA:DefinitionandSelectedPropertiesofGaussHypergeometricFunction-Subsection1-Equation2}, giving exactly \eqref{TheTrappingProbabilityandInfinitesimalGenerator-Section3-Equation12}, as required.
\end{proof}

\begin{figure}[h]
	\begin{subfigure}[b]{0.5\linewidth}
  		\includegraphics[scale=0.5]{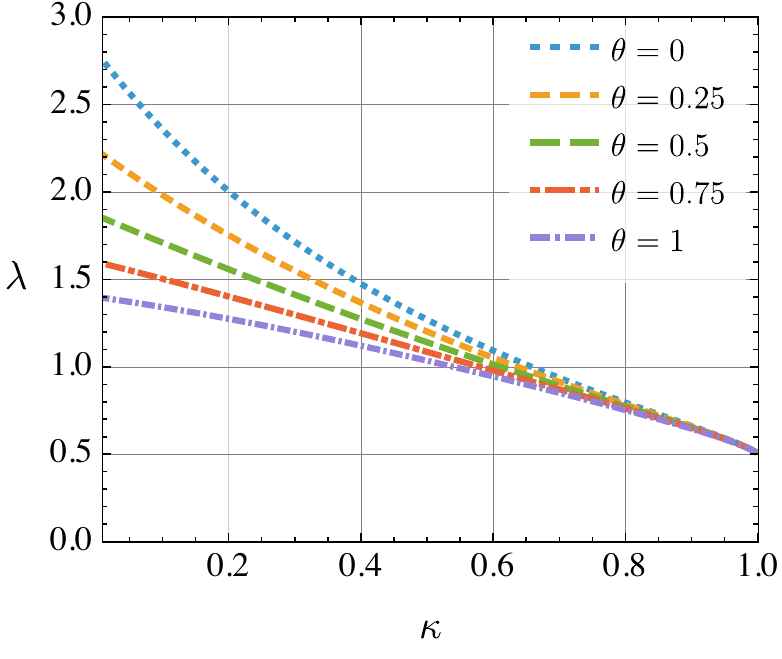}
		\caption{}
  		\label{TheTrappingProbabilityandInfinitesimalGenerator-Section3-Figure1-a}
	\end{subfigure}
	\begin{subfigure}[b]{0.5\linewidth}
  		\includegraphics[scale=0.5]{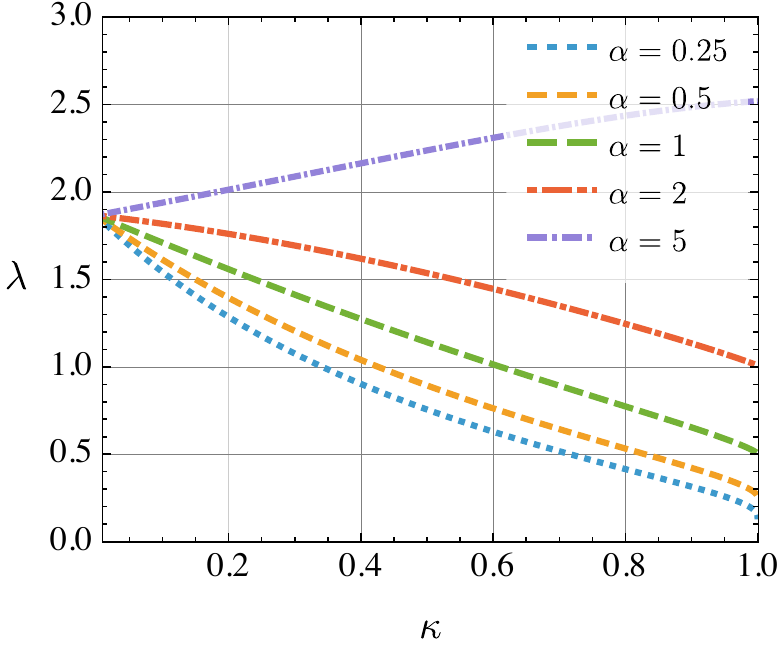}
		\caption{}
  		\label{TheTrappingProbabilityandInfinitesimalGenerator-Section3-Figure1-b}
	\end{subfigure}
	\caption{Upper boundary of the region defined by the net profit condition of \eqref{TheTrappingProbabilityandInfinitesimalGenerator-Section3-Equation12} in terms of $\kappa$ and $\lambda$ for $a = 0.1$, $b = 1.4$, $c = 0.4$ with (a) fixed $\alpha=1$ and different values of $\theta$ and (b) fixed $\theta=0.5$ and different values of $\alpha$. Note that $r =r(\kappa,\lambda)$ is given in Equation \eqref{TheCapitalModel-Section2-Equation10}.} \label{TheTrappingProbabilityandInfinitesimalGenerator-Section3-Figure1} 
\end{figure}

Condition \eqref{TheTrappingProbabilityandInfinitesimalGenerator-Section3-Equation12} reflects the competing effects of proportional insurance on the household's capital process. Insurance reduces the proportion of each loss retained by the household, while the associated premium payments reduce the effective capital growth rate $r$. Moreover, $r$ is not an independent parameter, but is related
to parameters modeling the household income growth, consumption and saving rates, as noted in \eqref{TheCapitalModel-Section2-Equation10}.
Therefore,  the net profit condition \eqref{TheTrappingProbabilityandInfinitesimalGenerator-Section3-Equation12} may be interpreted as defining the admissible combinations of $\lambda$ and $\kappa$ for fixed values of the remaining parameters. In particular, it implies that $r>0$, and hence automatically satisfies the requirement $b>p$.  The dependence of the net profit condition on the model parameters is shown in Figure \ref{TheTrappingProbabilityandInfinitesimalGenerator-Section3-Figure1}.

\begin{figure}[h]
	\begin{subfigure}[b]{0.5\linewidth}
  		\includegraphics[scale=0.6]{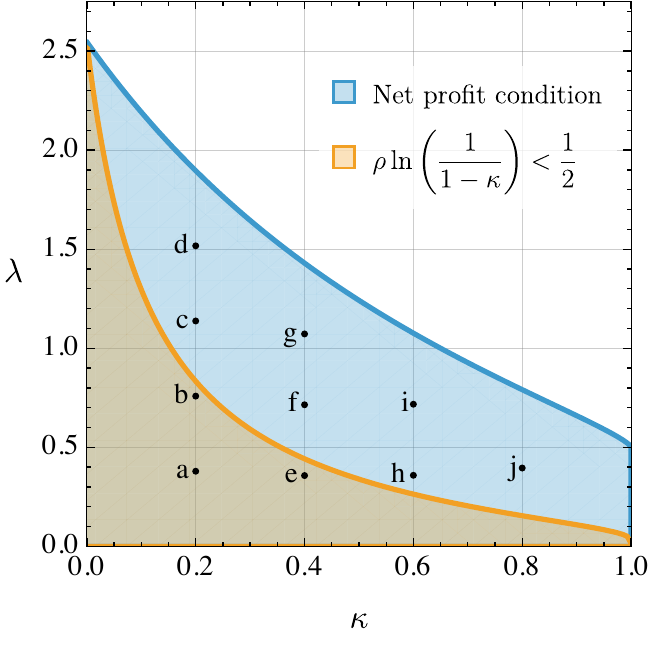}
		\caption{}
  		\label{fig:regions_kappa_lambda}
	\end{subfigure}
	\begin{subfigure}[b]{0.5\linewidth}
  		\includegraphics[scale=0.6]{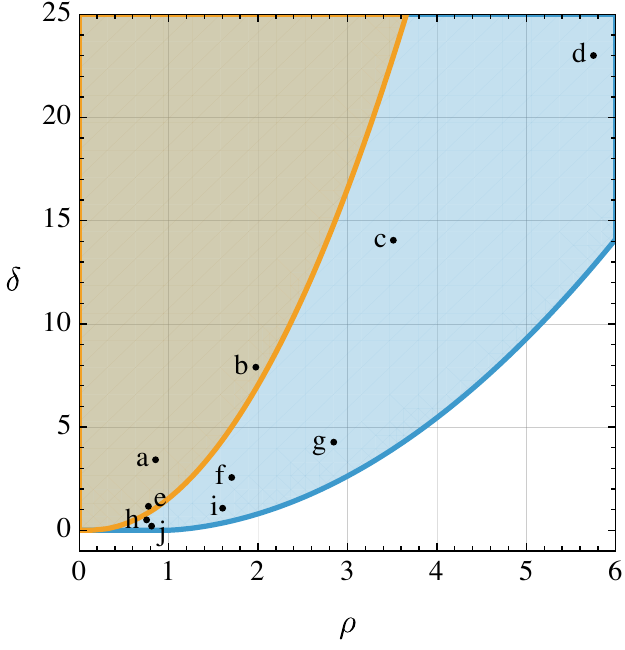}
		\caption{}
  		\label{fig:regions_rho_delta}
	\end{subfigure}
	    \caption{Regions in the $(\kappa,\lambda)$ plane corresponding to the various conditions of interest for an insured household under proportional losses with $Z_{i}\sim \text{Beta}(1,1)$. The parameters are set to $a = 0.1$, $b = 1.4$, $c = 0.4$ and $\theta = 0.1$. Note that the condition under which $h$ is monotone in Lemma \ref{TheTrappingProbabilityandInfinitesimalGenerator-Section3-Lemma1} implies the net profit condition \eqref{TheTrappingProbabilityandInfinitesimalGenerator-Section3-Equation17}. Here, $\rho = \frac{\lambda}{r}$ and $\delta = \rho\left(\frac{1-\kappa}{\kappa}\right)$, as defined in Remark \ref{AnalyticalApproachtoTrappingProbabilitiesviaInfinitesimalGenerators-Section4-Remark3}.}
    \label{fig:regions} 
\end{figure}

Figure \ref{TheTrappingProbabilityandInfinitesimalGenerator-Section3-Figure1-a} illustrates the dependence of the net profit condition on $\theta$ for $\alpha =1$, that is, when $Z_i$ is uniformly distributed on $[0,1]$. This case will be treated in detail in Section \ref{AnalyticalApproachtoTrappingProbabilitiesviaInfinitesimalGenerators-Section4-Subsection2} and for it, the net profit condition \eqref{TheTrappingProbabilityandInfinitesimalGenerator-Section3-Equation12} reduces to:

\begin{align}
    \frac{1}{\rho} = \frac{r}{\lambda} > 1+\frac{1-\kappa}{\kappa}\ln(1-\kappa).
    \label{TheTrappingProbabilityandInfinitesimalGenerator-Section3-Equation17}
\end{align}
Observe that lowering the loading factor $\theta$ increases the region in which certain trapping is prevented when remaining proportions are uniformly distributed. This is intuitive, since a lower value of $\theta$ corresponds to more affordable insurance coverage for households. Furthermore, note that the sensitivity of the constraint to the loading factor $\theta$ increases for decreasing $\kappa$ and thus increasing insurance coverage.

In contrast, Figure \ref{TheTrappingProbabilityandInfinitesimalGenerator-Section3-Figure1-b} varies $\alpha$ and shows that, as the shape parameter $\alpha$ increases, the region in which certain trapping is prevented becomes wider. This is due to the fact that higher values of $\alpha$ imply a larger expected proportion of capital remaining after a household experiences a loss. It is worth noting that the sensitivity of the constraint to the shape parameter $\alpha$ decreases for decreasing $\kappa$ and thus increasing insurance coverage. Indeed, decreasing $\kappa$ mitigates the impact of losses on the household\rq s capital (see, e.g., Figure \ref{TheCapitalModel-Section2-Figure1-b}). As a result, when insurance coverage is sufficiently high, differences in the shape parameter $\alpha$ have a progressively smaller effect on the household\rq s post-loss capital and, consequently, on the boundary of the net profit condition.

The net profit condition established here provides the parameter regime under which the analytical characterisation of the insured trapping probability in Section \ref{AnalyticalApproachtoTrappingProbabilitiesviaInfinitesimalGenerators-Section4-Subsection2} is meaningful. Further interactions among the various parameter restrictions are summarised in Figure \ref{fig:regions}. In particular, the net profit condition in the uninsured model corresponds to $\kappa=1$. Thus, as shown in Figure \ref{fig:regions_kappa_lambda}, in the absence of insurance coverage, the net profit condition is satisfied only for cases \textup{a}, \textup{e}, \textup{h}, and \textup{j}, whereas it is violated for cases \textup{b}, \textup{c}, \textup{d}, \textup{f}, \textup{g}, and \textup{i}. On the other hand, once insurance coverage is introduced, the net profit condition is satisfied for all cases considered, as all points lie within the blue region corresponding to the insured framework. Indeed, as it will be further discussed in Section \ref{Discussion-Section5}, it follows from \eqref{npc:rho} and \eqref{TheTrappingProbabilityandInfinitesimalGenerator-Section3-Equation17} that some households that were facing certain trapping without insurance, have a positive probability of avoiding trapping if insurance is acquired, even though the critical capital $x_0^*$ increases. For completeness, Figure \ref{fig:regions_rho_delta} also presents the regions of the net profit condition in the $(\rho, \delta)$ plane.

\section{Analytical Approach to Trapping Probabilities}
\label{AnalyticalApproachtoTrappingProbabilitiesviaInfinitesimalGenerators-Section4}

Having established the parameter regimes under which the trapping probability is non-trivial, we now turn to its analytical characterisation. Our objective in this section is to derive explicit and recursive representations of the trapping probability for the uninsured and insured capital models, respectively. 

The approach taken here differs from that of \cite{Article:Kovacevic2011}. Their analysis proceeds in two stages: an integral equation is first solved for the transition probability of the capital process, which is then used in a second integral equation for the trapping probability. In contrast,
we work directly with the trapping probability through the infinitesimal
generator of the capital process. This leads to a boundary value problem for the trapping probability itself,
allowing us to bypass the intermediate calculation of the transition
probabilities.

Our generator approach can be formulated using a classical result due to \cite{Article:Paulsen1997}, which shows that the trapping probability is a bounded function on $[x^{*}, \infty)$, twice continuously differentiable with bounded first derivative on $(x^{*}, \infty)$, and satisfies the following boundary value problem:

\begin{align}
    \mathcal{A}\left[\psi\right](x) = 0 \text{ for all } x \in (x^*, \infty)\quad  \text{and} \quad \psi(x^*)=1, 
    \label{TheTrappingProbabilityandInfinitesimalGenerator-Section3-Equation3}
\end{align}

where $\mathcal{A}$ is the infinitesimal generator of the stochastic process $\{X_t\}_{t\geq 0}$ given by either \eqref{InfinitesimalGeneratorNoInsurance} or \eqref{TheTrappingProbabilityandInfinitesimalGenerator-Section3-Equation18}. Furthermore, if the net profit condition holds (as detailed in Sections \ref{NetProfitConditionfortheUninsuredModel} and \ref{TheTrappingProbabilityandInfinitesimalGenerator-Section3-Subsection3}), we have
 $\lim_{x\rightarrow\infty}\psi(x)=0.$

The remainder of this section develops analytical solutions of this boundary
value problem. We begin with the uninsured model in Section
\ref{AnalyticalApproachtoTrappingProbabilitiesviaInfinitesimalGenerators-Section4-Subsection1},
where a closed-form expression for the trapping probability is available.
We then turn to the proportionally insured model in Section
\ref{AnalyticalApproachtoTrappingProbabilitiesviaInfinitesimalGenerators-Section4-Subsection2},
where the non-local structure of the resulting equation leads to the recursive
layer method developed in this paper. All proofs of the results in this section
are provided in Appendix \ref{AppendixB:MathematicalProofs}.

\subsection{Trapping Probability Under No Insurance Coverage} \label{AnalyticalApproachtoTrappingProbabilitiesviaInfinitesimalGenerators-Section4-Subsection1}
 We show that in the uninsured capital model, the boundary value problem admits a closed-form solution. Besides serving as a benchmark for the insured case, the explicit expression reveals the asymptotic behaviour of the trapping probability.

The following proposition presents the closed-form expression of the trapping probability for the uninsured setting when $Z_{i} \sim \text{Beta}(\alpha,1)$. In this case, $\psi$ satisfies:
\begin{align}
   \mathcal{A}[\psi](x)=r(x-x^{*})\psi'(x)-\lambda \psi(x)+ \lambda\int_{0}^1 \psi(x\cdot z) \alpha z^{\alpha-1}\rmd z = 0,\quad x \geq x^{*},
\label{TheTrappingProbabilityandInfinitesimalGenerator-Section3-Equation11}
\end{align}
with boundary conditions $\psi(x^*)=1$ and $\lim_{x\rightarrow\infty} \psi(x) = 0.$

\begin{proposition} 
\label{AnalyticalApproachtoTrappingProbabilitiesviaInfinitesimalGenerators-Section4-Proposition1}

Consider an uninsured household capital process as proposed in Definition \ref{TheCapitalModel-Section2-Definition1} with initial capital $x\geq x^{*}$, capital growth rate $r$, loss intensity $\lambda > 0$ and remaining proportions of capital $Z_{i}$ with distribution $\text{Beta}(\alpha,1)$. Assume the net profit condition $\rho =\tfrac{\lambda}{r} < \alpha$ in Proposition \ref{TheTrappingProbabilityandInfinitesimalGenerator-Section3-Proposition2} holds. Then, the trapping probability is given by

\begin{align}
\psi(x) = \frac{\Gamma(\alpha)}{\Gamma(\rho)\Gamma(\alpha-\rho)}
\left(\frac{x^*}{x}\right)^{\alpha-\rho} \int_0^1 t^{\alpha-\rho-1} \left(1 - \frac{x^*}{x}t\right)^{\rho-1} \rmd t.
\label{AnalyticalApproachtoTrappingProbabilitiesviaInfinitesimalGenerators-Section4-Equation1}
\end{align}

\end{proposition}

\begin{remark} \label{AnalyticalApproachtoTrappingProbabilitiesviaInfinitesimalGenerators-Section4-Remark1}

Although our primary interest is vulnerable non-poor households with capital levels relatively close to the poverty threshold, the asymptotic behavior of the trapping probability is useful for understanding the long-run effect of proportional losses and for comparison with classical additive-loss models. It can be obtained directly from the explicit form of the solution. Indeed, the integral term in \eqref{AnalyticalApproachtoTrappingProbabilitiesviaInfinitesimalGenerators-Section4-Equation1} satisfies
\begin{align}
    \lim_{x\rightarrow \infty} \int_0^1 t^{\alpha-\rho-1}  \left(1- t\frac{x^*}{x}\right)^{\rho-1} \rmd t = \int_0^1 t^{\alpha-\rho-1} \rmd t = \frac{1}{\alpha-\rho}.
\label{AnalyticalApproachtoTrappingProbabilitiesviaInfinitesimalGenerators-Section4-Equation11}
\end{align}

Thus 

\begin{align}
    \psi(x) \sim  \left(\frac{x^*}{x}\right)^{\alpha-\rho} \frac{\Gamma(\alpha)}{\Gamma(\rho)\Gamma(\alpha-\rho)} \frac{1}{\alpha-\rho}
    =  \left(\frac{x^*}{x}\right)^{\alpha-\rho} \frac{\Gamma(\alpha)}{\Gamma(\rho)\Gamma(\alpha-\rho+1)},
\label{AnalyticalApproachtoTrappingProbabilitiesviaInfinitesimalGenerators-Section4-Equation12}
\end{align}

as established in Equation (3.3.8) of \cite{Book:Henshaw2022}. Equation  \eqref{AnalyticalApproachtoTrappingProbabilitiesviaInfinitesimalGenerators-Section4-Equation12} shows that the trapping probability  decays algebraically with exponent $\alpha-\rho$. This differs with the exponential decay observed in the classical Cramér–Lundberg model with light-tailed claims (see, e.g., \cite{Book:Asmussen2010,Book:Mandjes2023}) and highlights a fundamental qualitative difference between the present capital process and the classical surplus process. 

\end{remark}

\begin{figure}[H]
	\begin{subfigure}[b]{0.5\linewidth}
 		\includegraphics[scale=0.6]{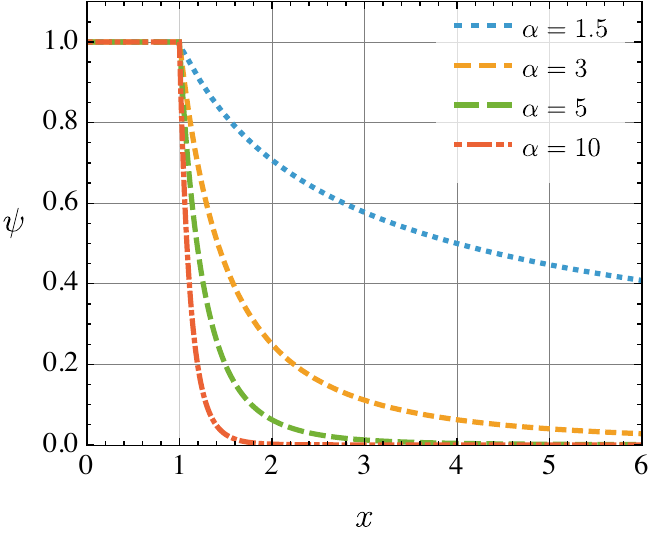}
		\caption{}
 		\label{AnalyticalApproachtoTrappingProbabilitiesviaInfinitesimalGenerators-Section4-Figure1-a}
	\end{subfigure}
	\begin{subfigure}[b]{0.5\linewidth}
 		\includegraphics[scale=0.6 ]{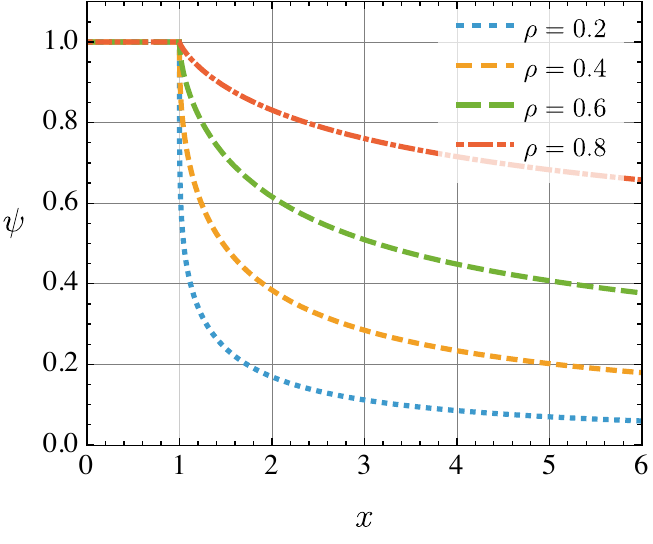}
		\caption{}    \label{AnalyticalApproachtoTrappingProbabilitiesviaInfinitesimalGenerators-Section4-Figure1-b}
	\end{subfigure}
	\caption{Trapping probability \eqref{AnalyticalApproachtoTrappingProbabilitiesviaInfinitesimalGenerators-Section4-Equation1} under no insurance for $Z_{i} \sim \text{Beta}(\alpha, 1)$, $a = 0.1$, $b = 1.4$, $c = 0.4$, $\lambda = 1$ and $x^{*} = 1$. (a) Fixed $\rho =1$ and varying $\alpha$ (b) Fixed $\alpha =1$ and varying $\rho$. The parameter values satisfy  the net profit condition $\rho < \alpha$ in Proposition \ref{TheTrappingProbabilityandInfinitesimalGenerator-Section3-Proposition2}.}
    \label{AnalyticalApproachtoTrappingProbabilitiesviaInfinitesimalGenerators-Section4-Figure1}
\end{figure}

Figure \ref{AnalyticalApproachtoTrappingProbabilitiesviaInfinitesimalGenerators-Section4-Figure1-a} illustrates the effect that the varying initial capital $x$ and shape parameter $\alpha$ have on the trapping probability determined in Proposition \ref{AnalyticalApproachtoTrappingProbabilitiesviaInfinitesimalGenerators-Section4-Proposition1}. As highlighted before, higher values of $\alpha$ increase the mean of the distribution of the remaining proportion of capital. Observation of a decreasing trapping probability with increasing $\alpha$ is therefore intuitive and aligns with the reduction in loss. 

Alternatively, Figure \ref{AnalyticalApproachtoTrappingProbabilitiesviaInfinitesimalGenerators-Section4-Figure1-b} presents the trapping probability \eqref{AnalyticalApproachtoTrappingProbabilitiesviaInfinitesimalGenerators-Section4-Equation1} for varying loss frequency $\lambda$ and fixed $\alpha=1$. In this case, the remaining proportions of capital are uniformly distributed, a scenario that is examined in Section \ref{AnalyticalApproachtoTrappingProbabilitiesviaInfinitesimalGenerators-Section4-Subsection2} for insured households. Note that the trapping probability tends to one as $\rho$ tends to $\alpha$ in line with the net profit condition found in Proposition \ref{TheTrappingProbabilityandInfinitesimalGenerator-Section3-Proposition2}. Increasing the frequency of loss events increases the trapping probability, as is to be expected.

Unlike the uninsured model, no closed-form solution is available under proportional insurance. We therefore develop a recursive analytical method for solving the corresponding boundary value problem.

\subsection{Trapping Probability Under Proportional Insurance Coverage} \label{AnalyticalApproachtoTrappingProbabilitiesviaInfinitesimalGenerators-Section4-Subsection2}

Before describing the method used to solve for the trapping probability in the proportional insured model, we establish a maximum principle for solutions of the equation $\mathcal{A}[h](x) =0$, where $\mathcal{A}$ is the infinitesimal generator of the insured capital process defined by \eqref{TheTrappingProbabilityandInfinitesimalGenerator-Section3-Equation18} in Section \ref{TheCapitalModel-Section2-Subsection2}. Under some conditions on the parameters $\rho$ and $\kappa$, we also show that these solutions are bounded.  Specifically, we have the following result:

\begin{lemma}
\label{TheTrappingProbabilityandInfinitesimalGenerator-Section3-Lemma1}
Assume that $h$ is a function with the following properties: (i) continuous on $[(1-\kappa)x^*,\infty)$; (ii) constant on $[(1-\kappa)x^*,x^*]$; (iii) differentiable on $(x^*,\infty)$; (iv) there exists $\epsilon > 0$ such that $h'(x) \neq 0$ for all $x \in (x^*,x^*+\epsilon)$; and (v) $\mathcal{A}[h](x) = 0$ for $x>x^*$. Then, if $G_Y$ is absolutely continuous with respect to the Lebesgue measure, $h$ is strictly monotone on $[x^*,\infty)$. If furthermore,
\begin{equation}\label{eq:bounded_cond}
    \rho \ln\left(\frac{1}{1-\kappa}\right)  < \frac12,
\end{equation}
then $h$ is bounded and thus $ \lim_{x\rightarrow\infty} h(x)$ exists and is finite.
\end{lemma}

In Section \ref{sec:LayerMethod}, our analysis relies on the assumptions of Lemma \ref{TheTrappingProbabilityandInfinitesimalGenerator-Section3-Lemma1}. In Lemma \ref{AnalyticalApproachtoTrappingProbabilitiesviaInfinitesimalGenerators-Section4-Lemma1}, we show that condition \eqref{eq:bounded_cond} implies that the net profit condition is satisfied. Consequently, the analysis in Section \ref{sec:LayerMethod} is restricted to the region defined by \eqref{TheTrappingProbabilityandInfinitesimalGenerator-Section3-Equation17}, thus excluding the cases in which households are trapped with probability one due to a failure of the net profit condition.

For the rest of the paper, we focus on the scenario in which $Z_i \sim \text{Beta}(1,1)$ in the case of proportional insurance, i.e., the remaining proportions of capital are uniformly distributed. In this case, the integral equation satisfied by the trapping probability, $\mathcal{A}[\psi](x) = 0$, where $\mathcal{A}$ is given by \eqref{TheTrappingProbabilityandInfinitesimalGenerator-Section3-Equation18}, can be replaced by an equivalent second order differential equation, albeit non-local as shown in Proposition \ref{AnalyticalApproachtoTrappingProbabilitiesviaInfinitesimalGenerators-Section4-Proposition2}. Despite its non-locality, the resulting equation can be solved using a layer method described in this section resulting in Proposition \ref{prop:layer_method}. This section finalises with the proof of the main result in this paper, Theorem \ref{maintheorem}, and details the construction of the fundamental solution set needed in the layer method. Although the case of uniform $Z_i$ serves as a natural benchmark in which no particular remaining proportion within $[0,1]$ is favored, 
a solution for general distributions would be desirable. However, if for example, $Z_i$ has a general Beta distribution, Equation \eqref{TheTrappingProbabilityandInfinitesimalGenerator-Section3-Equation18} cannot be written as a differential equation, and thus requires a different treatment from the one we propose.

From Section \ref{TheCapitalModel-Section2-Subsection2}, we know that the remaining proportion of capital after each loss event $Y_i$, is given, in the case of proportional insurance, by $Y_i = 1-\kappa(1-Z_i)$; see equation \eqref{eq:defYi}.  Since we are assuming that $Z_i$ is uniformly distributed on $[0,1]$, $Y_i$ is uniformly distributed on the interval $[1-\kappa, 1].$
Accordingly,  $\rmd G_{Y}(y) = {\rmd y}/{\kappa}$ and from Equation \eqref{TheTrappingProbabilityandInfinitesimalGenerator-Section3-Equation18}, the infinitesimal generator of the insured capital process can be rewritten as:

\begin{align}
\mathcal{A}[h](x) = r (x-x^{*}) h^{\prime}(x) - \lambda h(x) + \frac{\lambda}{\kappa x} \int_{(1-\kappa)x}^x h(y) \rmd y.
\label{AnalyticalApproachtoTrappingProbabilitiesviaInfinitesimalGenerators-Section4-Equation18}
\end{align}

\begin{remark}\label{AnalyticalApproachtoTrappingProbabilitiesviaInfinitesimalGenerators-Section4-Remark3}

Since capital is measured in units of the critical capital $x^{*}$, we henceforth set $x^{*}=1$. On the other hand, time is measured in units of $\tfrac{1}{\lambda}$. Recalling that  $\rho= \tfrac{\lambda}{r}$, we introduce the following non-dimensional quantity: $\delta:= \rho \left(\frac{1-\kappa}{\kappa}\right)$. From \eqref{TheCapitalModel-Section2-Equation10}, the parameters $\kappa$, $\lambda$, and  $r$ can be written in terms of $\rho$ and $\delta$ as follows:

\begin{align}
    \kappa(\rho,\delta) = \frac{\rho}{\rho + \delta}, \quad 
    \lambda(\rho,\delta) &=\frac{2 b c \rho (a-1)(\delta +\rho)}{\delta 
   ((a-1) c (\theta +1) \rho -2)-2 \rho},
   \quad \text{ and } \quad r(\rho,\delta)= \frac{\lambda(\rho,\delta)}{\rho}.
\label{AnalyticalApproachtoTrappingProbabilitiesviaInfinitesimalGenerators-Section4-Equation21}
\end{align}

In terms of $\rho$ and $\delta$ the net profit condition in Equation \eqref{TheTrappingProbabilityandInfinitesimalGenerator-Section3-Equation17}, reads
\begin{align}
    \rho - \delta \ln\left(1+\frac{\rho}{\delta}\right) < 1. \label{AnalyticalApproachtoTrappingProbabilitiesviaInfinitesimalGenerators-Section4-Equation51}
\end{align}
\end{remark} 

To find a solution to $\mathcal{A}[h](x) = 0$, where $h$ is continuous  on $[0,\infty)$ and twice differentiable on $(1,\infty)$, we manipulate \eqref{AnalyticalApproachtoTrappingProbabilitiesviaInfinitesimalGenerators-Section4-Equation18} to obtain a related differential equation. Specifically, we have that

\begin{align}
    \frac{1}{r} \frac{\rmd  (x\mathcal{A}[h](x))}{\rmd x} = 
    \mathcal{L}_\delta[h](x) - \delta h((1-\kappa)x),
    \label{AnalyticalApproachtoTrappingProbabilitiesviaInfinitesimalGenerators-Section4-Equation22}
\end{align}

where $\mathcal{L}_\delta$ is the  Sturm-Liouville operator defined by

\begin{align}
    \mathcal{L}_\delta[h](x) = x (x-1) h^{\prime \prime}(x) +[(2-\rho) x - 1] h^{ \prime}(x) + \delta h(x).
    \label{AnalyticalApproachtoTrappingProbabilitiesviaInfinitesimalGenerators-Section4-Equation23}
\end{align}

For ease of presentation a slight abuse of notation takes place.  The dependence on $\rho$ and $\delta$  of the infinitesimal generator $\mathcal{A}$ is suppressed.  However, we explicitly keep the dependence on $\delta$ of the differential operator $\mathcal{L}_\delta$.  This will be used later to relate solutions of these operators for different values of $\delta.$ In Proposition \ref{AnalyticalApproachtoTrappingProbabilitiesviaInfinitesimalGenerators-Section4-Proposition2} this abuse of notation results in stating a relation between the parameterised $\mathcal{L}_\delta$ with the not explicitly parameterised $\mathcal{A}$. 

\begin{proposition}\label{AnalyticalApproachtoTrappingProbabilitiesviaInfinitesimalGenerators-Section4-Proposition2}

Let $h$ be a continuous  function on $[0,\infty)$ that is constant on $[0,1]$ and twice continuously differentiable for $x>1$. Then, 
    
\begin{align}
        \mathcal{A}[h](x)= 0, \quad \text{for } x \geq 1,
        \label{AnalyticalApproachtoTrappingProbabilitiesviaInfinitesimalGenerators-Section4-Equation24}
\end{align}

if and only if,

\begin{align}
\mathcal{L}_\delta[h](x) = \delta h((1-\kappa)x), \quad \text{for } x \geq 1.
\label{AnalyticalApproachtoTrappingProbabilitiesviaInfinitesimalGenerators-Section4-Equation25}
\end{align}

\end{proposition}

Proposition \ref{AnalyticalApproachtoTrappingProbabilitiesviaInfinitesimalGenerators-Section4-Proposition2} establishes the equivalence between \eqref{AnalyticalApproachtoTrappingProbabilitiesviaInfinitesimalGenerators-Section4-Equation24} and \eqref{AnalyticalApproachtoTrappingProbabilitiesviaInfinitesimalGenerators-Section4-Equation25}. Therefore,  the introduction of proportional insurance results in a second order differential equation with multiplicative delay, which is notably different from the uninsured case discussed in Section \ref{AnalyticalApproachtoTrappingProbabilitiesviaInfinitesimalGenerators-Section4-Subsection1}. The second-order part of $\mathcal L_\delta$ is also degenerate at the poverty threshold $x=1$, since the coefficient $x(x-1)$ of $h''(x)$ vanishes there. This motivates expressing the local differential operator in the canonical speed-scale form used in the theory of one-dimensional diffusions. In particular, the speed and scale measures absorb the singular behavior at $x=1$ and provide the Sturm-Liouville structure needed to construct the corresponding Green's function and, subsequently, the recursive layer solution as documented in Section \ref{sec:LayerMethod}. To proceed, we write  the operator $\mathcal{L}_{\delta}$ in Sturm-Liouville form:

\begin{align}
    \mathcal{L}_{\delta}[h](x) = \frac{\rmd}{\rmd m}\frac{\rmd}{\rmd s}h(x) + \delta h(x),
\label{AnalyticalApproachtoTrappingProbabilitiesviaInfinitesimalGenerators-Section4-Equation26}
\end{align}

where the speed and scale measures are $    \rmd m(x) = \frac{\rmd x}{(x-1)^{\rho}}$ and $ \rmd s(x) = \frac{\rmd x}{x(x-1)^{1-\rho}},$ respectively.

\subsubsection{The Layer Method} \label{sec:LayerMethod}
We seek the trapping probability $\psi$ which satisfies \eqref{AnalyticalApproachtoTrappingProbabilitiesviaInfinitesimalGenerators-Section4-Equation24} or \eqref{AnalyticalApproachtoTrappingProbabilitiesviaInfinitesimalGenerators-Section4-Equation25} in Proposition \ref{AnalyticalApproachtoTrappingProbabilitiesviaInfinitesimalGenerators-Section4-Proposition2} with the boundary conditions $\psi(x) = 1$  for  $x \in [0,1] $ and $ \lim_{x\to\infty} \psi(x) = 0.$ The Ansatz for the trapping probability is:

\begin{align}
    \psi(x) = 1 - \frac{f(x)}{L},
    \label{eq:ansatz}
\end{align}

where the \textit{auxiliary function} $f$ satisfies  \eqref{AnalyticalApproachtoTrappingProbabilitiesviaInfinitesimalGenerators-Section4-Equation25} with boundary conditions 

\begin{align} f(x) = 0 \text{ for } x \in [0,1] \quad \text { and }
    \quad \exists ~L=  \lim_{x\to\infty} f(x) >0. \label{AnalyticalApproachtoTrappingProbabilitiesviaInfinitesimalGenerators-Section4-Equation30}
\end{align}

Observe that the function $\frac{1}{L}{f(x)}$ is the non-trapping probability, i.e., the probability that a household never falls below the poverty line. To obtain $f$, we solve the differential equation in each layer $\left[1/(1-\kappa)^{n-1},1/(1-\kappa)^{n}\right]:=[x_{n-1},x_n]$ of $[1,\infty)=\bigcup_{n=1}^\infty[x_{n-1},x_n]$ by iteratively using classical Sturm-Liouville theory. A schematic illustration of our  layer method is shown in Figure \ref{fig:layer-method-roadmap}.

\begin{figure}[H]
\centering

\resizebox{0.90\textwidth}{!}{
\begin{tikzpicture}[
    x=2.35cm,
    y=1cm,
    >=Stealth,
    every node/.style={font=\small},
    layer/.style={
        draw,
        rounded corners=3pt,
        minimum height=2.8cm,
        text width=4.2cm,
        align=left,
        inner sep=8pt
    }
]

\node[layer] (I0) at (1.0,3.2) {
    \centering
    \textbf{Initial Layer: $I_0=[0,1]$}\\[3pt]
    \par\medskip
    \raggedright
    The auxiliary function $f(x)$ is known:
    \[
        f(x)=0 \text{ for all } x\in[0,1].
    \]
};

\node[layer] (I1) at (3.1,3.2) {
    \centering
    \textbf{Layer 1: $I_1=(1, x_1]$}\\[3pt]
    \par\medskip
    \raggedright
    If $x\in I_1$, then
        $x(1-\kappa)\in I_0$,
    where $f$ is already known.\\[3pt]

    Solve the ODE on $I_1$ to obtain $f_1(x)$.
};

\node[layer] (I2) at (5.2,3.2) {
    \centering
    \textbf{Layer 2: $I_2=(x_1, x_2]$}\\[3pt]
    \par\medskip
    \raggedright
    If $x\in I_2$, then
        $x(1-\kappa)\in I_1$,
    where $f$ is already known.\\[3pt]

    Solve the ODE on $I_2$ to obtain $f_2(x)$.
};

\draw[thick,->] (0,0) -- (7.2,0)
    node[right] {$x$};

\foreach \xx/\lab in {
    0/{0},
    2.05/{1},
    4.15/{x_1},
    6.25/{x_2}
}{
    \fill (\xx,0) circle (2pt);
    \node[below=4pt] at (\xx,0) {$\lab$};
}

\draw[dashed] (0,0.15) -- (0,1.55);
\draw[dashed] (2.05,0.15) -- (2.05,1.55);
\draw[dashed] (4.15,0.15) -- (4.15,1.55);
\draw[dashed] (6.25,0.15) -- (6.25,1.55);

\node[below=0.55cm] at (1,0)
    {$I_0=[0,1]$};

\node[below=0.55cm] at (3,0)
    {$I_1=(1,x_1]$};

\node[below=0.55cm] at (5,0)
    {$I_2=(x_1,x_2]$};

\fill (3.25,-1.75) circle (2pt);

\node[above=2pt] at (3.25,-1.75)
    {$x\in I_1$};

\draw[dashed,->]
    (3.25,-1.75)
    .. controls (3,-2.2) and (2,-2.5) ..
    (1.45,-3.35);

\node[below=2pt] at (1.45,-3.35)
    {$x(1-\kappa)\in I_0$};

\fill (5.25,-1.75) circle (2pt);

\node[above=2pt] at (5.25,-1.75)
    {$x\in I_2$};

\draw[dashed,->]
    (5.25,-1.75)
    .. controls (4.8,-2.2) and (3.9,-2.5) ..
    (3.45,-3.35);

\node[below=2pt] at (3.45,-3.35)
    {$x(1-\kappa)\in I_1$};

\node[align=center] at (6.75,3.2)
    {\textbf{Continue}\\
     \textbf{recursively}};

\draw[dashed,->,thick]
    (6.3,2.4) -- (7.25,2.4);

\node[right] at (6.25,-0.02)
    {$\cdots$};

\end{tikzpicture}%
}

\caption{
The positive real number line is partitioned into successive layers
$I_0=[0,1]$, $I_1=(1,x_1]$, $I_2=(x_1,x_2]$, and so on.
On each new layer, the delayed argument $x(1-\kappa)$ lies in a
previously constructed layer, allowing the corresponding differential
equation to be solved recursively.}
\label{fig:layer-method-roadmap}

\end{figure}

\begin{proposition} \label{prop:layer_method}

Let $u$ and $w$ be linearly independent solutions to the homogeneous problem  $\mathcal{L}_\delta[\cdot](x) = 0,$ with $u(1)= 1$, $w(1)=0$, and $\tfrac{\rmd w}{\rmd s}(1^+)>0$. Define the Green\rq s function, $G$, for the initial value problem and the corresponding Wronskian, $W$, by 

\begin{align}
        G(x,x') = \frac{w(x)u(x') - u(x)w(x')}{W} \quad \text{ and } \quad 
        W = u(x) \frac{\rmd w}{\rmd s}(x) - w(x) \frac{\rmd u}{\rmd s}(x) >0,
        \label{AnalyticalApproachtoTrappingProbabilitiesviaInfinitesimalGenerators-Section4-Equation32}
\end{align}

respectively. For $n=1,2,\dots$, let $x_n = \frac{1}{(1-\kappa)^n}$ and recursively define the sequence of functions $\{z_n\}_{n=1}^{\infty}$ by
\begin{align}
    z_1(x) &=  \begin{cases}
        0, & 0<x<1,\\
        w(x), & 1 \leq x,
    \end{cases}\\
        z_n(x) &= \begin{cases}
        0, & 0<x<x_{n-1},\\
        \displaystyle{\delta \int_{x_{n-1}}^x  \! z_{n-1}((1-\kappa)x') ~ G(x,x') ~ \rmd m(x')}, &  x_{n-1} \leq x.
    \end{cases}
    \label{AnalyticalApproachtoTrappingProbabilitiesviaInfinitesimalGenerators-Section4-Equation36}
\end{align}

Furthermore, define  $\{f_n\}_{n=1}^{\infty}$ as

\begin{align}
    f_n(x) &= z_1(x) + \dots + z_n(x), \quad 0 \leq x \leq x_n.
    \label{AnalyticalApproachtoTrappingProbabilitiesviaInfinitesimalGenerators-Section4-Equation37}
\end{align}

Then, the point-wise limit $f = \lim_{n\to \infty} f_{n}$ solves \eqref{AnalyticalApproachtoTrappingProbabilitiesviaInfinitesimalGenerators-Section4-Equation24} and \eqref{AnalyticalApproachtoTrappingProbabilitiesviaInfinitesimalGenerators-Section4-Equation25} on $[1,\infty)$. Moreover, if $0<\rho<1$ or $\rho \ln(1/(1-\kappa)) <1/2,$ $f$ is bounded and has a finite limit $L = \lim_{x \to \infty} f(x)>0$.

\end{proposition}

We make two additional observations about the conditions under which Proposition \ref{prop:layer_method} ensures boundedness of $f$. First, as Lemma \ref{AnalyticalApproachtoTrappingProbabilitiesviaInfinitesimalGenerators-Section4-Lemma1} states, they guarantee that the net profit condition in \eqref{AnalyticalApproachtoTrappingProbabilitiesviaInfinitesimalGenerators-Section4-Equation51} holds; see Figure \ref{fig:regions}. This ensures, by Proposition \ref{TheTrappingProbabilityandInfinitesimalGenerator-Section3-Proposition3}, that $\lim_{x \to \infty} \psi(x) = 0$, which is consistent with the behavior of the Ansatz in \eqref{eq:ansatz}.

\begin{lemma}\label{AnalyticalApproachtoTrappingProbabilitiesviaInfinitesimalGenerators-Section4-Lemma1}

    If $0<\rho<1$ or  $\rho \ln(1/(1-\kappa)) < 1/2$ then the net profit condition $\frac{1}{\rho} > 1+\frac{1-\kappa}{\kappa}\ln(1-\kappa)$ holds for all $\kappa \in (0,1)$.
    
\end{lemma}

Second, the condition if $0<\rho<1$ or  $\rho \ln(1/(1-\kappa)) < 1/2$ seems to be more restrictive than necessary for the existence of the limit $L$. Indeed, the numerical simulations presented in Section \ref{Discussion-Section5} include cases of parameters outside the region identified in the statement of the proposition in which boundedness can be established. Specifically, the values of $\kappa$ and $\lambda$ corresponding to the points labeled \textup{c, d, f, g} and $\textup{i}$ in Figure \ref{fig:regions}, fall outside of this region. Nevertheless, the numerical calculations in Section \ref{Discussion-Section5} illustrate that the auxiliary function $f$ is bounded and thus it can be used to determine the trapping probability via the Ansatz in \eqref{eq:ansatz}.

Theorem \ref{maintheorem} gives the construction of the trapping probability:

\begin{theorem}
\label{maintheorem}
Let $f$ be constructed as in Proposition \ref{prop:layer_method} and suppose that  $\psi(x')$ is known for some $x'>1$. Then, the trapping probability is given by

\begin{align}
    \psi(x) = \begin{cases}
        1, & 0\leq x \leq 1,\\
        1- \frac{1-\psi(x')}{f(x')} f(x), & 1 \leq x.
    \end{cases}
    \label{eq:psi_final}
\end{align}

\end{theorem}

Thus, the layer construction determines the functional form of the trapping probability up to a single normalizing quantity, \(L\) (or equivalently, the trapping probability at one reference point). To summarise, computing the trapping probability by the layer method and Theorem \ref{maintheorem}, requires linearly independent solutions $\{u,w\}$ as in Proposition \ref{prop:layer_method}, and an estimate for $\psi(x')$ at some $x' > 1$, or equivalently, an estimate of the limiting value $L$ of the auxiliary function $f$. 

\begin{figure}[]
	\begin{subfigure}[b]{0.5\linewidth}
 		\includegraphics[scale=0.65]{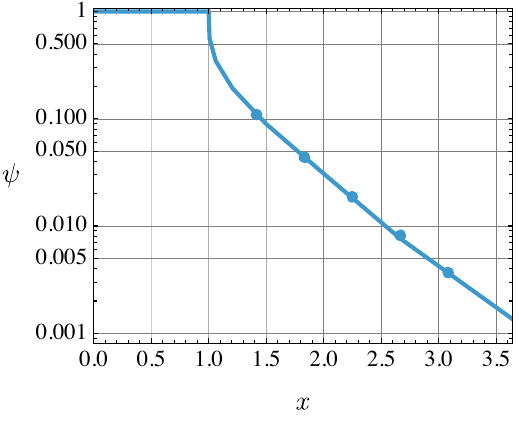}
	\end{subfigure}
	\begin{subfigure}[b]{0.5\linewidth}
 		\includegraphics[scale=0.65]{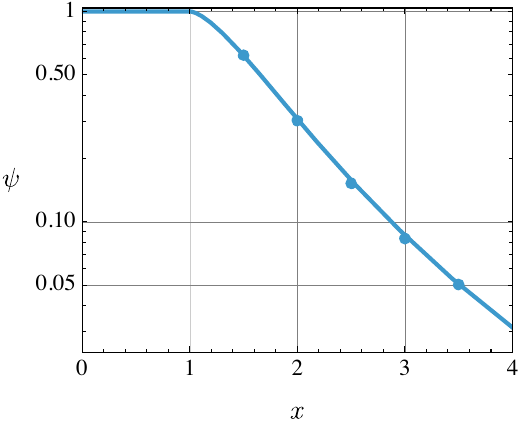}
	\end{subfigure}
	\caption{
    Comparison between simulated values of the trapping probability and our method. The left graph shows $\psi(x)$ for the case $(\rho,\delta) = (0.2,0.1)$ which corresponds to $(\kappa,\lambda) = (0.67,0.1)$. The right graph corresponds to the case labeled with the letter f in Figure \ref{fig:regions}. Solid lines show the function $\psi(x)$ computed via the layer method described in Section \ref{sec:LayerMethod}. Dots show the estimation of $\psi(x)$ obtained by directly simulating $250,000$ trajectories of the process $X_t$ with $X_0=x$, and counting the number of realisations that are trapped by $t=50$. These estimates are performed independently from those used to estimate $L$.}
    \label{fig:psi_final01}
\end{figure}

For completeness,  we give a full  construction for the functions $\{u,w\}$ in Appendix \ref{Trapping Probability Under Proportional Insurance Coverage - Analytic Solutions}, which are then used to compute $f$ via the layer method. This is illustrated in Figure \ref{fig:illustration}. The method we use to estimate $L$ is as follows: we consider five values of $x' \in [1,x_1)$, where $f(x')$ equals the analytically known $w(x')$. For each, we estimate the value of $\psi(x')$ by simulating $20,000$ realisations of the process $X_t$ with $X_0 = x'$ and counting the number of trapped trajectories by time $t = 100$. The value of $L$ is then taken by averaging the expression \eqref{eq:findL} over the five cases. This method is used for the computation of $\psi$ in Figures \ref{fig:psi_final01} and \ref{fig:psi_compare} used in the comparisons, graphs and general discussion presented in Section \ref{Discussion-Section5}.

\section{Model Comparisons and the Impact of Insurance} \label{Discussion-Section5}

The analytical results of Section \ref{AnalyticalApproachtoTrappingProbabilitiesviaInfinitesimalGenerators-Section4} allow us to examine two features of the model that are of particular interest. First, we compare the asymptotic behavior of the trapping probability under proportional losses with that obtained under the additive loss model of \cite{Article:Flores-Contro2024}. This comparison highlights a fundamental difference between the two loss mechanisms. Second, we use the recursive solution developed in Section \ref{AnalyticalApproachtoTrappingProbabilitiesviaInfinitesimalGenerators-Section4-Subsection2} to examine how proportional insurance affects the trapping probability across different initial capital levels. In particular, we identify the conditions under which this coverage is beneficial, in the sense that it reduces the trapping probability relative to the case without insurance.

\subsection{Additive versus Multiplicative Losses}
\label{Discussion-Section5-Subsection1}

We first compare the asymptotic decay of the uninsured household trapping probability under proportional (multiplicative) losses with that obtained under the random-valued (additive) losses. For exponentially distributed additive losses, the uninsured trapping probability for ${x}\geq x^*$ is given by (see Equation (3.3) in \cite{Article:Flores-Contro2024}):
\begin{align}
\psi^{{\scaleto{\text{ {\fontfamily{qcr}\selectfont EXP}}}{2.5pt}}}(x)=\frac{\Gamma\left(\rho;\mu(x-x^*)\right)}{\Gamma\left(\rho\right)},
\label{AnalyticalApproachtoTrappingProbabilitiesviaInfinitesimalGenerators-Section4-Equation15}
\end{align}
where $\rho = \tfrac{\lambda}{r}$ and $\Gamma(a;z):=\int_{z}^{\infty}e^{-t}t^{a-1}\rmd t$ is the upper incomplete gamma function and $\mu$ is the exponential loss parameter.
The asymptotic behaviour as $x\rightarrow \infty$ of the trapping probability in \eqref{AnalyticalApproachtoTrappingProbabilitiesviaInfinitesimalGenerators-Section4-Equation15} is given by
\begin{align}
\psi^{{\scaleto{\text{ {\fontfamily{qcr}\selectfont EXP}}}{2.5pt}}}(x) \sim \mu^{\rho-1}(x-x^*)^{\rho-1}e^{-\mu(x-x^*)}(1+O(|\mu(x-x^*)|^{-1})),
\label{AnalyticalApproachtoTrappingProbabilitiesviaInfinitesimalGenerators-Section4-Equation16}
\end{align}
so the limiting behavior as $x \to \infty$ of the ratio of \eqref{AnalyticalApproachtoTrappingProbabilitiesviaInfinitesimalGenerators-Section4-Equation16} to \eqref{AnalyticalApproachtoTrappingProbabilitiesviaInfinitesimalGenerators-Section4-Equation12} is
\begin{align}
\frac{\psi^{{\scaleto{\text{ {\fontfamily{qcr}\selectfont EXP}}}{2.5pt}}}(x)}{\psi(x)} \sim {Cx^{\alpha-\rho}(x-x^{*})^{\rho-1}e^{-\mu({x}-x^*)}}(1+O(|\mu({x}-x^*)|^{-1})),
\label{AnalyticalApproachtoTrappingProbabilitiesviaInfinitesimalGenerators-Section4-Equation17}
\end{align}
for some constant $C$. Thus, the trapping probability under additive exponential losses decays strictly faster  than under multiplicative losses, with the severity of this difference dependent on the parameters of the loss distributions. This reflects the scale dependence of the two loss mechanisms. Under additive losses, a random loss becomes relatively less important as initial capital increases. Under multiplicative losses, by contrast, the distribution of the fraction of capital lost does not change with the capital level. Consequently, even households with large initial capital remain exposed to losses representing a substantial proportion of their wealth. 

\begin{figure}[H]
	\begin{subfigure}[b]{0.5\linewidth}
 		\includegraphics[scale=0.65]{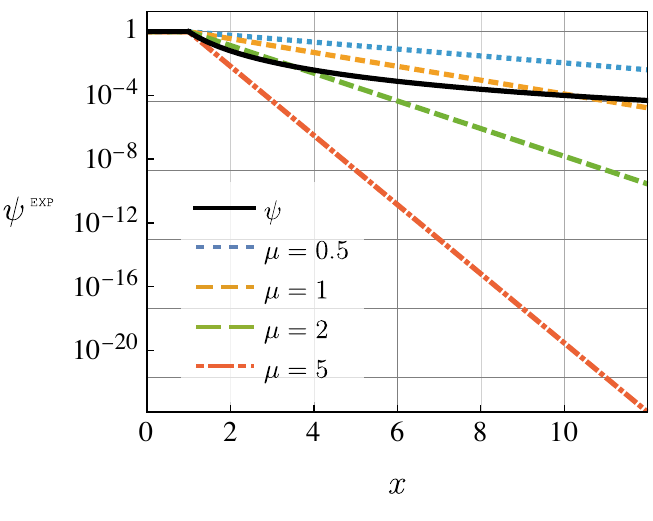}
		\caption{}
        \label{fig:psi_exp_mus}
	\end{subfigure}
	\begin{subfigure}[b]{0.5\linewidth}
 		\includegraphics[scale=0.65]{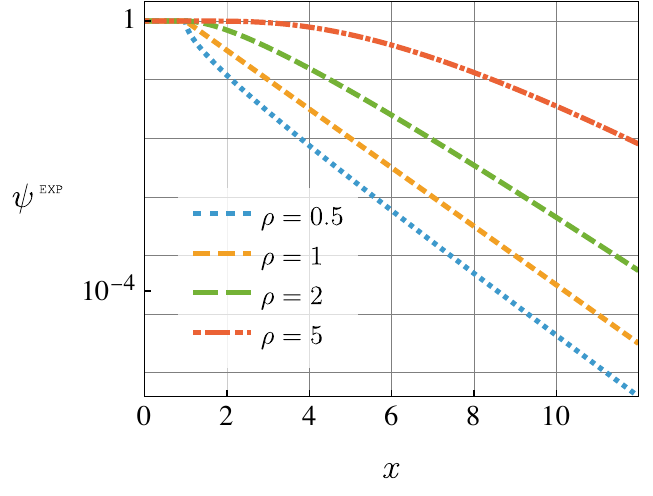}
		\caption{}    \label{fig:psi_exp_rhos}
	\end{subfigure}
	\caption{(a) Comparison of the trapping probability $\psi^{{\scaleto{\text{ {\fontfamily{qcr}\selectfont EXP}}}{2.5pt}}}(x)$ 
    given in \eqref{AnalyticalApproachtoTrappingProbabilitiesviaInfinitesimalGenerators-Section4-Equation15} for random-valued losses with $Z_{i}\sim\text{Exponential}(\mu)$ and different values of $\mu$, against the trapping probability $\psi(x)$ in \eqref{AnalyticalApproachtoTrappingProbabilitiesviaInfinitesimalGenerators-Section4-Equation1} for proportional losses with $Z_{i}\sim\text{Beta}(5,1)$ (black solid line). Here, the parameters are given by $a = 0.1$, $b = 1.4$, $c = 0.4$, $\rho = 1$ and $x^{*} = 1$ (b) The trapping probability $\psi^{{\scaleto{\text{ {\fontfamily{qcr}\selectfont EXP}}}{2.5pt}}}(x)$ in \eqref{AnalyticalApproachtoTrappingProbabilitiesviaInfinitesimalGenerators-Section4-Equation15} for different values of $\rho$.}
    \label{fig:psi_exp}
\end{figure}

A comparison of the decay of the trapping probability under proportional losses against that of random-valued losses is provided in Figure \ref{fig:psi_exp}. Here, the slower rate of decay under proportional losses is clearly observable. Figure \ref{fig:psi_exp_mus} compares the trapping probabilities under the case in which the remaining proportions of capital are Beta$(5, 1)-$distributed given by \eqref{AnalyticalApproachtoTrappingProbabilitiesviaInfinitesimalGenerators-Section4-Equation1} and exponentially distributed additive losses given by \eqref{AnalyticalApproachtoTrappingProbabilitiesviaInfinitesimalGenerators-Section4-Equation15} for different values of $\mu$. For random-valued claim sizes with an expected value of 0.5 ($\mu=2$) the trapping probability is greater than for proportional losses for the most vulnerable (households with initial capital levels close to the poverty line), however, as capital increases the trapping probability under proportional losses exceeds the random-valued case. If the expected claim size increases to one ($\mu=1$) the trapping probability for proportional losses is significantly lower than in the random-valued case for a higher portion of the most vulnerable. Compared to the mean loss associated with $\text{Beta}(5,1)-$distributed remaining proportions, an expected claim size of one is low with respect to high levels of initial capital. Figure \ref{fig:psi_exp_rhos} illustrates the trapping probability under the random-valued loss model \eqref{AnalyticalApproachtoTrappingProbabilitiesviaInfinitesimalGenerators-Section4-Equation15} for different values of $\rho$. Note that the net profit condition holds here, so the trapping probability decreases to zero as the initial capital increases.

\subsection{Impact of Proportional Insurance on Trapping Probability} \label{Discussion-Section5-Subsection2}

\begin{figure}[p]
    \centering
    \includegraphics[scale=0.465]{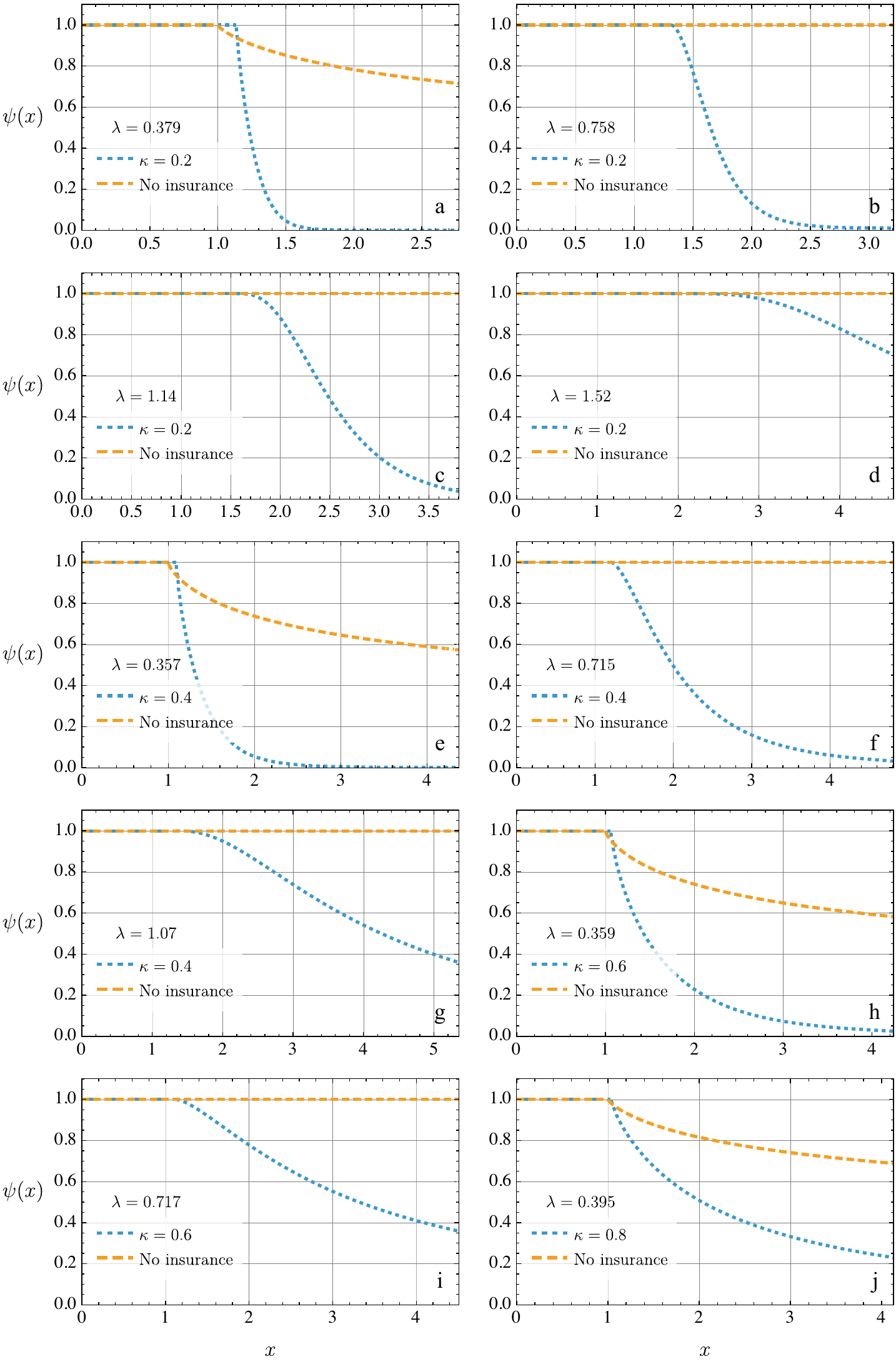}
    \caption{Comparison of the trapping probabilities for uninsured and insured households (with proportional insurance coverage) under proportional losses with $Z_{i}\sim\text{Beta}(1,1)$. The parameters are set to $a = 0.1$, $b = 1.4$, $c = 0.4$, $\lambda = 1$ and $x_{0}^{*} = 1$. Each subplot corresponds to one of the cases illustrated in Figure \ref{fig:regions} (refer to the label in the bottom-right corner of each subplot). The trapping probabilities for uninsured households (orange dashed curves) and insured households (blue dotted curves) are computed using \eqref{AnalyticalApproachtoTrappingProbabilitiesviaInfinitesimalGenerators-Section4-Equation1} and \eqref{eq:psi_final}, respectively.}
    \label{fig:psi_compare}
\end{figure}

Figure \ref{fig:psi_compare} compares the trapping probabilities of uninsured and insured households for different values of $\lambda$ and $\kappa$. The probabilities for uninsured households are computed using the closed-form expression \eqref{AnalyticalApproachtoTrappingProbabilitiesviaInfinitesimalGenerators-Section4-Equation1}, whereas those for insured households are obtained from the recursive solution presented in Section \ref{AnalyticalApproachtoTrappingProbabilitiesviaInfinitesimalGenerators-Section4-Subsection2}. In line with the developments in Section \ref{AnalyticalApproachtoTrappingProbabilitiesviaInfinitesimalGenerators-Section4-Subsection2}, we consider the case in which the remaining proportions of capital are $\text{Beta}(1,1)-$distributed.

A particularly important effect of insurance is worth highlighting. For several parameter regimes considered in Figure \ref{fig:psi_compare}, specifically plots labeled \textup{b, c, d, f, g} and \textup{i}, the uninsured model violates the net profit condition $\rho<1$ determined in Proposition \ref{TheTrappingProbabilityandInfinitesimalGenerator-Section3-Proposition2}. Hence, trapping occurs with certainty for all uninsured households, regardless of their initial capital level, yielding a constant trapping probability of one. As explained in Section \ref{TheCapitalModel-Section2-Subsection2}, insurance modifies both the retained-loss distribution and the capital growth rate. Consequently, it can shift a household for which escape from the poverty trap is impossible into a parameter regime satisfying the insured net profit condition, thereby creating a positive possibility of escape. In other words, proportional insurance fundamentally changes the qualitative behaviour of the process: whereas trapping is certain in the uninsured model, insurance reduces the trapping probability to a value strictly below one, with this probability decreasing as the initial capital increases.

Furthermore, we also observe in Figure \ref{fig:psi_compare} that for some households with initial capital levels barely above the poverty threshold  $x^{*}_{0}$, purchasing proportional insurance coverage is detrimental as their trapping probabilities increase after acquiring this protection (see, e.g., plots labeled \textup{a, e, h} and \textup{j}). The increase for this particular group is consistent with previous findings in \cite{Article:Kovacevic2011} and \cite{Article:Flores-Contro2024}, and arises because paying the insurance premium reduces a household\rq s income generation rate, thereby raising the critical capital threshold to a higher level, $x^{*} \geq x^{*}_{0}$, as explained in Section \ref{TheCapitalModel-Section2-Subsection2}. 

Despite being disadvantageous for some of the most vulnerable households, Figure \ref{fig:psi_compare} shows that proportional insurance coverage is beneficial for the great majority of households. Specifically, the figure suggests the existence of a threshold $x_{\bullet} > x_{0}^{*}$ such that, for $x_{0}^{*} < x < x_{\bullet}$, the uninsured household has the lower trapping probability. 
Hence, vulnerable households with capital levels just above the poverty line may require support to make insurance beneficial for them (see, for instance, \cite{Article:Flores-Contro2024}, which introduces two different frameworks for offering subsidies to households, thereby reducing the impact of premium payments on their income-generating capacity).

The comparisons presented in Sections \ref{Discussion-Section5-Subsection1} and \ref{Discussion-Section5-Subsection2} reveal two distinct effects of proportional losses and proportional insurance. Multiplicative losses fundamentally alter the tail behavior of the trapping probability, producing algebraic rather than exponential decay. Insurance, meanwhile, 
has two distinguishable effects.  On the one hand, it shows that households that can afford insurance can avoid the certainty of trapping since the NPC \eqref{TheTrappingProbabilityandInfinitesimalGenerator-Section3-Equation12} holds for a larger set of parameters than the case of no-insurance
\eqref{npc:rho}. Insurance, however, does not uniformly reduce trapping risk. Indeed, households with capital barely above the trapping probability would not be able to afford the purchase of an insurance and thus remain under the poverty level.    
In particular, while premium payments may increase the likelihood of falling into the poverty trap for households with initial capital levels that are close to the poverty line, insurance can eliminate the certainty of trapping for households with higher initial capital.

\section{Conclusion} \label{Conclusion-Section6}

This article studies the capital process introduced in \cite{Article:Kovacevic2011}, where households are susceptible to losses that are proportional to their accumulated capital. Under the assumption of proportional losses, households at all capital levels are exposed to events that can result in substantial reductions in capital. This feature is particularly relevant for vulnerable non-poor households, i.e., those living just above the poverty line, but also distinguishes the model from classical risk processes with additive losses, since even households with relatively high levels of capital remain exposed to losses representing a significant proportion of their wealth.

Focusing on the probability that a household ever falls below the poverty line, referred to as the trapping probability, we consider two cases: (i) households with no insurance coverage and (ii) households with proportional insurance coverage. We first develop an alternative characterisation of the trapping time through an embedded discrete-time process. This representation leads naturally to a net profit condition, analogous to that in classical ruin theory, which identifies the parameter regimes under which the infinite-time trapping probability is non-trivial. In particular, when the expected logarithmic accumulation is negative, trapping occurs with probability one, whereas under the net profit condition the trapping probability converges to zero as the initial capital tends to infinity.

For households with no insurance coverage, explicit infinite-time trapping probabilities are obtained when the remaining proportions of capital are $\mathrm{Beta}(\alpha,1)-$distributed. The resulting expression also provides the asymptotic behaviour of the trapping probability, which decays algebraically as the initial capital increases. This differs from the exponential decay arising in classical additive-loss models and, in particular, rules out exponential decay for the proportional-loss model considered here. Comparison with the additive (random-valued) loss framework of \cite{Article:Flores-Contro2024} further illustrates the slower decay produced by multiplicative losses.
 
Incorporating proportional insurance coverage modifies both the capital growth rate and the distribution of the remaining proportion of capital after a loss. The corresponding infinitesimal generator therefore contains non-local terms, and the resulting integro-differential equation does not admit the same explicit treatment as in the uninsured case. To address this, we transform the problem into a delay differential equation and develop a recursive layer method, based on Sturm--Liouville theory and the special functions that arise in this setting. This provides a constructive solution of the insured problem and allows the trapping probability to be evaluated numerically.

Comparisons between trapping probabilities under no insurance coverage and proportional insurance coverage show that there are parameter regimes in which an uninsured household faces certain eventual trapping because the net profit condition is violated, while the corresponding insured household has a trapping probability strictly below one. This highlights the importance of insurance as a risk mitigation tool, as it can turn certain trapping into a positive possibility of escape for households that would otherwise be trapped with certainty. Our analyses also show that for some households with initial capital levels just above the poverty line, insurance and the associated premium payments can increase the probability of falling below the poverty line, consistent with the findings of earlier studies. Nevertheless, our comparisons indicate that
proportional insurance coverage reduces the trapping probability over a broad range of higher initial capital levels. 

There are several natural directions in which the present model could be extended. In particular, the proportion of insurance coverage is fixed, premiums are calculated using the expected value principle, and the insured analysis is developed for uniformly distributed remaining proportions of capital. We also focus only on the event that a household falls below the poverty line, rather than on its subsequent behaviour after trapping. Future work could consider endogenous insurance decisions, premium subsidies, and more general proportional-loss distributions. Even under the present assumptions, however, the model helps identify how premium payments and proportional loss protection interact to affect the long-run probability of poverty trapping.

\section*{Funding}  \label{Funding-Section} 

The work of K.H. and C.C. was supported by Engineering and Physical Sciences Research Council (EPSRC) [grant number EP/W522399/1]; and the EPSRC and ESRC Centre for Doctoral Training on Quantification and Management of Risk and Uncertainty in Complex Systems Environments [grant number EP/L015927/1]. José Miguel Flores-Contró gratefully acknowledges funding from the FWO and F.R.S.-FNRS under the Excellence of Science (EOS) programme, project ASTeRISK (40007517).

\section*{Acknowledgements}  \label{Acknowledgements-Section}

E.T. thanks the colleagues in the Department of Mathematical Sciences at the University of Liverpool for their hospitality during his Fall 2023 visit, when part of this work was conducted. 


\section*{Appendices}
\addcontentsline{toc}{section}{Appendices}
\renewcommand{\thesubsection}{\Alph{subsection}}
\setcounter{subsection}{0}

\numberwithin{equation}{subsubsection}

\subsection{The Gauss Hypergeometric Function}\label{AppendixA:DefinitionandSelectedPropertiesofGaussHypergeometricFunction}

In this section, we collect the definition and selected properties of the Gauss hypergeometric function that are used throughout the article. The material presented here is not intended to be exhaustive; rather, it provides the specific results and identities required for our analysis.

For a comprehensive treatment of hypergeometric functions, including their relations, transformation formulas, and domains of convergence, see \cite{Book:Abramowitz1964} and \cite{Book:Kristensson2010}.

\subsubsection{Definition} \label{AppendixA:DefinitionandSelectedPropertiesofGaussHypergeometricFunction-Subsection1}

The Gauss hypergeometric function is defined as: 

\begin{align}
    { }_{2} F_{1}(a, b ; c ; z)=\sum_{n=0}^{\infty} \frac{(a)_{n}(b)_{n}}{(c)_{n}} \frac{z^{n}}{n !},
    \label{AppendixA:DefinitionandSelectedPropertiesofGaussHypergeometricFunction-Subsection1-Equation1}
\end{align}

where $(a)_{n}=\frac{\Gamma(a+n)}{\Gamma(a)}$ denotes the Pochhammer symbol \citep{Book:Seaborn1991}. If $\textrm{Re}(c) > \textrm{Re}(b) >0$ and $z\in \mathbb{C}$, $z\notin [1,\infty)$ the Gauss hypergeometric function has the integral representation (see, for instance, \cite[p. 558]{Book:Abramowitz1964}):

\begin{align}
    _2F_1 (a,b;c;z)=\frac{\Gamma(c)}{\Gamma(b)\Gamma(c-b)}\int_0^1 t^{b-1}(1-t)^{c-b-1}(1-tz)^{-a}\,\rmd t. \label{AppendixA:DefinitionandSelectedPropertiesofGaussHypergeometricFunction-Subsection1-Equation2}
\end{align}

\subsubsection{Selected Properties} \label{AppendixA:DefinitionandSelectedPropertiesofGaussHypergeometricFunction-Subsection2}

A number of transformation formulas for $~\!_2F_1\left(a,b;c;z\right)$ can be derived. We present below those that are used in this manuscript.

\begin{align}
    &~\!_2F_1\left(a,b;c;z\right)=\frac{\Gamma(c)\Gamma(c-a-b)}{\Gamma(c-a)\Gamma(c-b)} ~\!_2F_1\left(a,b;a+b-c+1;1-z\right)\\
    &+(1-z)^{c-a-b}\frac{\Gamma(c)\Gamma(a+b-c)}{\Gamma(a)\Gamma(b)}~\!_2F_1\left(c-a,c-b;c-a-b+1;1-z\right) \quad |\arg{(1-z)}|<\pi.\\
    \label{AppendixA:DefinitionandSelectedPropertiesofGaussHypergeometricFunction-Subsection1-Equation3}
\end{align}

\begin{align}
    ~\!_2F_1\left(a,b;c;z\right)=\left(1 - z\right)^{c-a-b}~\!_2F_1\left(c-a,c-b;c;z\right)
    \label{AppendixA:DefinitionandSelectedPropertiesofGaussHypergeometricFunction-Subsection1-Equation4}
\end{align}

In particular, \eqref{AppendixA:DefinitionandSelectedPropertiesofGaussHypergeometricFunction-Subsection1-Equation4} is also known as Euler\rq s transformation.

\subsection{Mathematical Proofs}\label{AppendixB:MathematicalProofs}
In this section, we present the proofs of all the results in Section \ref{AnalyticalApproachtoTrappingProbabilitiesviaInfinitesimalGenerators-Section4}.

\subsubsection{Proof of Proposition \ref{AnalyticalApproachtoTrappingProbabilitiesviaInfinitesimalGenerators-Section4-Proposition1}}
\label{ProofofProp4.1}
\begin{proof}

The trapping probability satisfies $\mathcal{A}\left[\psi\right]\left(x\right) = 0$ where $\mathcal{A}$ is given by \eqref{TheTrappingProbabilityandInfinitesimalGenerator-Section3-Equation11}. This yields

\begin{align}
(x-x^*) \psi'(x) =\rho \psi(x) -\rho \int_0^1 \psi(x\cdot z) \rmd z^\alpha.
\label{AnalyticalApproachtoTrappingProbabilitiesviaInfinitesimalGenerators-Section4-Equation2}
\end{align}

Since 

\begin{align}
\frac{\rmd}{\rmd x} \int_0^1 \psi(x\cdot z)  \rmd z^\alpha =
\frac{\rmd}{\rmd x} \left(\frac{\alpha}{x^{\alpha}} \int_0^x \psi(t)   t^{\alpha -1}  \rmd t\right) = \frac{\alpha}{x}\left[ \psi(x) -   \int_0^1 \psi(x\cdot z)  \rmd z^\alpha\right],
\label{AnalyticalApproachtoTrappingProbabilitiesviaInfinitesimalGenerators-Section4-Equation3}
\end{align}

we have that 

\begin{align}
\rho \frac{\rmd}{\rmd x} \int_0^1 \psi(x\cdot z)  \rmd z^\alpha   
= \frac{\alpha}{x} (x-x^*) \psi'(x).
\label{AnalyticalApproachtoTrappingProbabilitiesviaInfinitesimalGenerators-Section4-Equation4}
\end{align}

Differentiation of \eqref{AnalyticalApproachtoTrappingProbabilitiesviaInfinitesimalGenerators-Section4-Equation2} leads to

\begin{align}
(x-x^*) \psi'' + \psi' = \rho \psi' - \frac{\alpha}{x} (x-x^*) \psi'.
\label{AnalyticalApproachtoTrappingProbabilitiesviaInfinitesimalGenerators-Section4-Equation5}
\end{align}

Upon rearranging, we arrive at

\begin{align}
 0= x(x-x^*) \psi'' + \left[(1+\alpha-\rho) x-\alpha x^*\right] \psi'.
\label{AnalyticalApproachtoTrappingProbabilitiesviaInfinitesimalGenerators-Section4-Equation6}
\end{align}

Since \eqref{AnalyticalApproachtoTrappingProbabilitiesviaInfinitesimalGenerators-Section4-Equation6} is first order in $\psi'$, separation of variables leads to

\begin{align}
\frac{\psi''}{\psi'} = - \left( \frac{\alpha}{x} + \frac{1-\rho}{x-x^*} \right),
\label{AnalyticalApproachtoTrappingProbabilitiesviaInfinitesimalGenerators-Section4-Equation7}
\end{align}

so that $\psi'(x) = C x^{-\alpha}(x-x^*)^{\rho-1}$ for some positive constant $C$. Note that $\psi'$ is integrable at infinity since $\rho<\alpha$.  So integrating on the interval $[x,\infty)$ and using the fact that $\lim_{x\rightarrow\infty} \psi(x) =0,$ we get

\begin{align}
\psi(x) = C \int_x^\infty v^{-\alpha} (v-x^*)^{\rho-1} \rmd v = C \int_x^\infty v^{-\alpha+\rho} \left( 1 - \frac{x^*}{v}\right)^{\rho-1} \frac{\rmd v}{v}.
\label{AnalyticalApproachtoTrappingProbabilitiesviaInfinitesimalGenerators-Section4-Equation8}
\end{align}

With the substitution $v=x/t$, we obtain

\begin{align}
\psi(x) = C \int_0^1 \left(\frac{x}{t}\right)^{-\alpha+\rho} \left(1-\frac{x^*}{x} t\right)^{\rho-1} \frac{\rmd t}{t} = \frac{C}{x^{\alpha-\rho}} \int_0^1 t^{\alpha-\rho-1}  \left(1-\frac{x^*}{x} t\right)^{\rho-1} \rmd t.
\label{AnalyticalApproachtoTrappingProbabilitiesviaInfinitesimalGenerators-Section4-Equation9}
\end{align}

The boundary condition at $x^*$ gives 

\begin{align}
C = \frac{(x^*)^{\alpha-\rho} }{ \int_0^1 t^{\alpha-\rho-1}  \left(1- t\right)^{\rho-1} dt} = (x^*)^{\alpha-\rho} \frac{\Gamma(\alpha)}{\Gamma(\rho)\Gamma(\alpha-\rho)},
\label{AnalyticalApproachtoTrappingProbabilitiesviaInfinitesimalGenerators-Section4-Equation10}
\end{align}

as claimed. Note that the integral representation of the trapping probability can be identified as a Gauss hypergeometric function as shown in Proposition 3.3.3 of \cite{Book:Henshaw2022}. Using \eqref{AppendixA:DefinitionandSelectedPropertiesofGaussHypergeometricFunction-Subsection1-Equation2}, we can rewrite the integral term as 

\begin{align}
\int_0^1 t^{\alpha-\rho-1} \left(1 - \frac{x^*}{x}t\right)^{\rho-1} \rmd t=\frac{\Gamma(\alpha-\rho) \Gamma(1)}{\Gamma (\alpha-\rho+1)} {}_2F_1\left(1-\rho,\alpha-\rho; \alpha-\rho+1; \frac{x^*}{x}\right).
\label{AnalyticalApproachtoTrappingProbabilitiesviaInfinitesimalGenerators-Section4-Equation13}
\end{align}

This yields 
\begin{align}
\psi(x)&= \frac{\Gamma(\alpha)}{\Gamma(\rho)\Gamma(\alpha-\rho)} \left(\frac{x^*}{x}\right)^{\alpha-\rho}\frac{1}{\alpha-\rho} {}_2F_1\left(1-\rho, \alpha-\rho;\alpha-\rho+1; \frac{x^*}{x}\right).
\label{AnalyticalApproachtoTrappingProbabilitiesviaInfinitesimalGenerators-Section4-Equation14}
\end{align}

which is exactly Equation (3.3.17) in \cite{Book:Henshaw2022}.
\end{proof}
\subsubsection{Proof of Lemma \ref{TheTrappingProbabilityandInfinitesimalGenerator-Section3-Lemma1}} \label{ProofofLemma3.1}
\begin{proof}For $x>x^*$, rewrite $\mathcal{A}[h](x)=0$ as
\begin{equation}
    h'(x)
    = \frac{\rho}{x-x^*}
     \int_{1-\kappa}^{1}
    \left(h(x)-h(x \cdot y)\right)\rmd G_Y(y).
    \label{eq:Ah-rewritten}
\end{equation}
Since $h$ is continuous and $G_Y$ is absolutely continuous, the right-hand side is continuous as a function of $x$. Hence $h'$ is continuous on $(x^*,\infty)$.

Assumption (\textit{iv}) ensures that $h$ does not have an accumulation of critical points on a neighborhood to the right of $x^*$, and thus we can define
$$x_c = \inf\{x>x^*: h'(x)=0\} \geq x^* + \epsilon.$$
If $x_c < \infty$, then $|h'(x)|>0$ for all $x^*<x<x_c$; and since $h'$ is continuous on $(x^*,\infty)$, we also have $h'(x_c)=0$. Thus, evaluating $\mathcal{A}[h](x_c)=0$ gives
\begin{align}
    0
    &=
    \lambda h(x_c)
    -
    \lambda\int_{1-\kappa}^{1} h(x_c \cdot y)\rmd G_Y(y) =
    \lambda\mathbb{E}\left[h(x_c)-h(x_c \cdot Y)\right].
    \label{TheTrappingProbabilityandInfinitesimalGenerator-Section3-Equation19}
\end{align}
Since $h$ is assumed constant on $[(1-\kappa)x^*,x^*]$ and strictly monotone on $(x^*,x_c)$, and $x_c \cdot Y<x_c$ almost surely, the random variable inside the expectation is either positive or negative with probability one, which contradicts \eqref{TheTrappingProbabilityandInfinitesimalGenerator-Section3-Equation19}. Hence $x_c=\infty$, and $h$ is strictly monotone on $[x^*,\infty)$.
 
We now suppose $h$ is strictly increasing and prove that condition \eqref{eq:bounded_cond} implies the boundedness of $h$. The proof in the case of decreasing $h$ is similar. 

We write $\mathcal{A}[h](x)=0$ as 
$$h'(x) = \frac{\rho}{x-x^*}  \mathbb{E}\left[h(x) - h(x \cdot Y)\right],$$
and use layers $x_n = \frac{x^*}{(1-\kappa)^n}$  as in Section \ref{sec:LayerMethod}. For $n\geq 2$, integrating from $x_{n-1}$ to $x_{n}$, we define
\begin{align}
    \Delta_n  \equiv h(x_{n}) - h(x_{n-1}) = \rho  \int_{x_{n-1}}^{x_{n}} \frac{\mathbb{E}\left[h(x) - h(x\cdot Y)\right]}{x-x^*}  \rmd x.
    \label{TheTrappingProbabilityandInfinitesimalGenerator-Section3-Equation20}
\end{align}
Now, since $h$ is increasing and $(1-\kappa) < Y < 1$ almost surely, one has that for $x_{n-1}< x < x_{n}$, 
\begin{align}
    h(x) - h(x\cdot Y) \leq h(x_{n}) - h(x_{n-2}) = \Delta_n  + \Delta_{n-1},
    \label{eq:h_Deltan}
\end{align}
with probability one. We thus define for $n \geq 2$,
\begin{align}
    \gamma_n =  \rho   \int_{x_{n-1}}^{x_{n}} \frac{1}{x-x^*} \rmd x 
=  \rho  \ln\left(\frac{1-(1-\kappa)^{n}}{1-(1-\kappa)^{n-1}} \frac{1}{1-\kappa}\right).
    \label{eq:def_gamman}
\end{align}
Then, combining \eqref{eq:h_Deltan} and \eqref{eq:def_gamman},
\begin{equation}\label{eq:Dn_gamman1}
    \Delta_n \leq \gamma_n (\Delta_n + \Delta_{n-1}), \quad n=2,3,\dots
\end{equation}

Note that 
\begin{equation}
    \lim_{n \to \infty} \gamma_n =  \rho \ln \left(\frac{1}{1-\kappa}\right) < \frac{1}{2},
\end{equation}
by hypothesis. Then, we can take $n_0$ large enough so that $\gamma_n < 1$ for all $n\geq n_0$, and write \eqref{eq:Dn_gamman1} as

\begin{align}
\Delta_n \leq \frac{\gamma_n}{1-\gamma_n} \Delta_{n-1} \leq p_n \Delta_{n_0}, \quad p_n:=  \left(\prod_{j=n_0+1}^n \frac{\gamma_j}{1-\gamma_j}\right), \quad n > n_0.
\label{eq:def_pn}
\end{align}

Let $N>n_0$, and sum over $n_0 + 1 \leq n\leq N$ to obtain

\begin{align}
h(x_N) - h(x_{n_0}) \leq \Delta_{n_0} ~ \sum_{n=n_0+1}^{N} p_n.
\label{TheTrappingProbabilityandInfinitesimalGenerator-Section3-Equation25}
\end{align}

By the ratio test, the series $\sum_n p_n$ converges if
$\lim_{n\rightarrow\infty} \frac{\gamma_n}{1 - \gamma_n} < 1$ 
which is equivalent to the assumption $\lim_{n\rightarrow\infty} \gamma_n< 1/2.$
Thus $\lim_{N\rightarrow\infty} h(x_N) <\infty$, which implies that $h$ is bounded.   The existence of the limit of $h(x)$ as $x$ tends to $\infty$ follows from the fact that $h$ is bounded and increasing.
\end{proof}
 \subsubsection{Proof of Proposition \ref{AnalyticalApproachtoTrappingProbabilitiesviaInfinitesimalGenerators-Section4-Proposition2}
 }
 \begin{proof}
Clearly, \eqref{AnalyticalApproachtoTrappingProbabilitiesviaInfinitesimalGenerators-Section4-Equation24} implies \eqref{AnalyticalApproachtoTrappingProbabilitiesviaInfinitesimalGenerators-Section4-Equation25}. Now, suppose that $h$ satisfies \eqref{AnalyticalApproachtoTrappingProbabilitiesviaInfinitesimalGenerators-Section4-Equation25}, and define $A(x): = \mathcal{A}[h](x)$.  Then $\tfrac{\rmd x A(x)}{\rmd x} = 0$ so $A(x) = C/x$ for some $C \in \mathbb{R}$. Evaluating \eqref{AnalyticalApproachtoTrappingProbabilitiesviaInfinitesimalGenerators-Section4-Equation18} as $x \downarrow 1$ gives 
\begin{equation}
\label{equivalence}
A(1) = \lim_{x\downarrow 1}  (x-1)\frac{\rmd h}{\rmd x} - \lambda h(1) + \lambda h(0).
\end{equation}
Since $h(x)=\tilde{a}$ on $[0,1]$, $h(x) = \tilde{a} + \tilde{h}(x)$, where $\tilde{h}$ satisfies, for $x>1$, 
\begin{equation}
    \label{homogeneous}
\mathcal{L}_\delta[\tilde{h}](x)=0.
\end{equation}
In Proposition \ref{prop:uv_real}  and Proposition \ref{prop:uv_complex} we identify a fundamental solution set for the operator $\mathcal{L}_\delta$ so that any solution of \eqref{homogeneous} has the property that for any $\rho>0,$
$
\lim_{x\downarrow 1}  (x-1)^{1-\rho}\frac{\rmd h}{\rmd x} >0$.  Consequently, the limit term in \eqref{equivalence} vanishes.  Using that $h$ is constant on $[0,1]$, we 
obtain $C = A(1) = 0$ and thus $\mathcal{A}[h](x) = 0$ for all $x \geq 1$, as claimed.
\end{proof}
\subsubsection{Proof of Lemma \ref{AnalyticalApproachtoTrappingProbabilitiesviaInfinitesimalGenerators-Section4-Lemma1}} \label{ProofofLemma4.1}

\begin{proof}
We need to show that, for $\kappa \in (0,1)$,
\begin{equation}\label{eq:lemmaBound_NPC}
    0<\rho<1 \quad \text{or}  \quad 
    \rho \ln\left(\frac{1}{1-\kappa}\right) < \frac12 \quad \Rightarrow \quad
     \frac{1}{\rho} > 1 + \frac{1-\kappa}{\kappa}\ln(1-\kappa).
\end{equation}
Clearly, if $0<\rho<1$ the implication follows. For the second case, note that it suffices to show that 
\begin{equation}
    2 \ln\left(\frac{1}{1-\kappa}\right) >  1 + \frac{1-\kappa}{\kappa}\ln(1-\kappa),
\end{equation}
which can be equivalently written as
\begin{equation}\label{eq:nu>0}
    \nu(\kappa):= -\ln(1-\kappa) - \frac{\kappa}{\kappa + 1} >0.
\end{equation}
To establish \eqref{eq:nu>0} for $\kappa \in (0,1)$, it is enough to note that $\nu(0) = 0$ and $\nu'(\kappa) = \frac{\kappa(\kappa+3)}{(1+\kappa)^2(1-\kappa)} > 0$, so $\nu(\kappa)$ is strictly increasing, and thus positive, for $\kappa \in (0,1)$.
\end{proof}

\subsubsection{Proof of Proposition \ref{prop:layer_method}}
\label{ProofofProposition4.3}

\begin{proof}
Before we provide the proof, several observations are in order. First, we assume that $\{u,w\}$ is a set of linearly independent, solutions to the homogeneous problem  \eqref{AnalyticalApproachtoTrappingProbabilitiesviaInfinitesimalGenerators-Section4-Equation31} with boundary conditions $u(1)= 1$, $w(1)=0$, and $\tfrac{\rmd w}{\rmd s}|_{1^+}>0$. We will later show in Appendix \ref{Trapping Probability Under Proportional Insurance Coverage - Analytic Solutions} how to construct such solutions. Second, we have defined the Wronskian using derivatives with respect to the scale measure $s$ to avoid singularities at $x=1$.  With this definition, $W(x)$ is constant as shown in \cite{Book:Ito2006}, and positive since $W(1^+)=\tfrac{\rmd w}{\rmd s}(1^+) >0$. Third, by properties of the Green\rq s function, for any $g$ sufficiently smooth and any $\xi \geq 1$, the function $h$ given by 

\begin{align}
    h(x) = \int_\xi^x g(x') G(x,x') \rmd m(x') \quad \text{solves} \quad
    \mathcal{L}_\delta [h](x) = g(x) \text{ for } x \in [\xi,\infty),
    \label{AnalyticalApproachtoTrappingProbabilitiesviaInfinitesimalGenerators-Section4-Equation33}
\end{align}

with initial condition $h(\xi) = h'(\xi) = 0$.
In the first layer $\left[1,\frac{1}{1-\kappa}\right]$, the function $f$ that we seek must solve the homogeneous problem 
\begin{align}
\mathcal{L}_\delta[f](x) = 0, \quad \text{ for } \quad x \in \left(1, \frac{1}{1-\kappa}\right)
\label{AnalyticalApproachtoTrappingProbabilitiesviaInfinitesimalGenerators-Section4-Equation31},
\end{align} 
so we take 

\begin{align}
    f(x) = \begin{cases}
        0, & 0<x<1,\\
        w(x), & 1 \leq x \leq \frac{1}{1-\kappa }.
    \end{cases}
    \label{AnalyticalApproachtoTrappingProbabilitiesviaInfinitesimalGenerators-Section4-Equation34}
\end{align}

Proceeding to the second layer, i.e., for $x \in \left[1/(1-\kappa),1/(1-\kappa)^{2}\right]$, our solution  satisfies $\mathcal{L}_\delta[f](x) = \delta w((1-\kappa)x)$. Thus, we utilise the Green\rq s function to obtain $f$ in that layer as given by:

\begin{align}
    f(x) = w(x) + \delta \int_{\frac{1}{1-\kappa }}^x w((1-\kappa)x') G(x,x') \rmd m(x'), \quad \text{ for } \frac{1}{1-\kappa } \leq x \leq \frac{1}{(1-\kappa)^2}.
    \label{AnalyticalApproachtoTrappingProbabilitiesviaInfinitesimalGenerators-Section4-Equation35}
\end{align}

Sequentially, by adding integrals with respect to the Green\rq s function, we find solutions on each layer, as detailed by the following arguments.  

Each $z_n$ is, by construction and  \eqref{AnalyticalApproachtoTrappingProbabilitiesviaInfinitesimalGenerators-Section4-Equation33}, a solution to

\begin{align}
    \mathcal{L}_{\delta}[z_n] = \delta z_{n-1}((1-\kappa) x), \quad x \geq x_{n-1}, \quad 
    z_n(x_{n-1})= 0. 
    \label{AnalyticalApproachtoTrappingProbabilitiesviaInfinitesimalGenerators-Section4-Equation38}
\end{align}

Let $f_n$ be as in \eqref{AnalyticalApproachtoTrappingProbabilitiesviaInfinitesimalGenerators-Section4-Equation37} and note that, if $x \leq x_{n-1}$, then $f_{n-1}(x) = f_n(x)$. Now, for $x \in [0,x_n]$, we have the following,

\begin{align}
    \mathcal{L}_{\delta}[f_n](x) &=  \mathcal{L}_{\delta}[z_1](x)  + \mathcal{L}_{\delta}[z_2](x)+ \dots + \mathcal{L}_{\delta}[z_{n}](x) \nonumber\\
    &= \delta z_1((1-\kappa)x) +\delta z_2((1-\kappa)x)+ \dots + \delta z_{n-1}((1-\kappa)x) \nonumber\\
    &= \delta f_{n-1}((1-\kappa)x) \nonumber \\
    &= \delta f_n((1-\kappa)x). 
    \label{AnalyticalApproachtoTrappingProbabilitiesviaInfinitesimalGenerators-Section4-Equation39}
\end{align}

Each function $f_n$ is constant on $[0,1]$, continuous $[0,x_n]$, twice-differentiable on $(1,x_n]$, and by Proposition \ref{AnalyticalApproachtoTrappingProbabilitiesviaInfinitesimalGenerators-Section4-Proposition2}, solves $\mathcal{A}[f_n] =0 $.  Moreover, from the definition of the Wronskian in \eqref{AnalyticalApproachtoTrappingProbabilitiesviaInfinitesimalGenerators-Section4-Equation32}, we have $\frac{\rmd f_n}{\rmd s}(1^+) = \frac{\rmd w}{\rmd s}(1^+) = W> 0$. Therefore, there exists $\epsilon>0$ such that $f_n'(x) \neq 0$ for all $x \in (1,1+\epsilon)$. The point-wise limit $f = \lim_{n\to \infty} f_n$ inherits all these properties, and by Lemma  \ref{TheTrappingProbabilityandInfinitesimalGenerator-Section3-Lemma1} is a positive and strictly increasing solution to $\mathcal{A}[f]=0$ on $(1,\infty)$.

If $\rho\ln(1/(1-\kappa)) <1/2$, the boundedness of $f$ follows from Lemma \ref{TheTrappingProbabilityandInfinitesimalGenerator-Section3-Lemma1}.  To establish that $f$ is bounded in the case $\rho<1$, write Equation \eqref{AnalyticalApproachtoTrappingProbabilitiesviaInfinitesimalGenerators-Section4-Equation39} as

\begin{align}
    \frac{\rmd}{\rmd m}\frac{\rmd f_n}{\rmd s}(x) + \delta f_n(x) = \delta f_n((1-\kappa)x), \quad 1<x\leq x_n, 
    \label{AnalyticalApproachtoTrappingProbabilitiesviaInfinitesimalGenerators-Section4-Equation40}
\end{align}

and integrate with respect to $\rmd m$ and $\rmd s$ to get

\begin{align}
    f_n(x) = \frac{\rmd f_n}{\rmd s}(1^+) \int_1^x \rmd s(x') 
    + \delta \int_1^x \int_1^{x'} (f_n((1-\kappa)x'') - f_n(x'')) \rmd m(x'') \rmd s(x').
    \label{AnalyticalApproachtoTrappingProbabilitiesviaInfinitesimalGenerators-Section4-Equation41}
\end{align}

Since $f_{n}$ is increasing, the second integral on the right side of Equation \eqref{AnalyticalApproachtoTrappingProbabilitiesviaInfinitesimalGenerators-Section4-Equation41} is negative and we can bound $f_{n}(x), x \in [1,x_n]$ uniformly in $n$ by

\begin{align}
    f_n(x) < \frac{\rmd f_n}{\rmd s}(1^+) \int_1^\infty \rmd s(x') = W \int_1^\infty \frac{1}{x'(x'-1)^{1-\rho}} \rmd x' = W \pi \csc(\pi \rho).
    \label{AnalyticalApproachtoTrappingProbabilitiesviaInfinitesimalGenerators-Section4-Equation42}
\end{align}

Since on $[1, x_n],$ $f(x)=f_n(x)$, $f$ is bounded on $[1,\infty)$, as claimed. 
\end{proof}

\subsubsection{
Proof of Theorem \ref{maintheorem}
}
\begin{proof}
Proposition \ref{prop:layer_method} establishes the existence of a monotone increasing function $f$ satisfying $\mathcal{L}_\delta[f](x) =\delta f((1-\kappa)x)$ for $x>1$, $f(1) = 0$ and such that there exists $L=\lim_{x\rightarrow\infty} f(x)>0$. Use the Ansatz \eqref{eq:ansatz} to define $\psi(x) = 1 - \frac{f(x)}{L}$. Then $\psi$ solves \eqref{AnalyticalApproachtoTrappingProbabilitiesviaInfinitesimalGenerators-Section4-Equation24} and \eqref{AnalyticalApproachtoTrappingProbabilitiesviaInfinitesimalGenerators-Section4-Equation25} in Proposition \ref{AnalyticalApproachtoTrappingProbabilitiesviaInfinitesimalGenerators-Section4-Proposition2} along with the boundary conditions in \eqref{AnalyticalApproachtoTrappingProbabilitiesviaInfinitesimalGenerators-Section4-Equation30}. The function $\psi$ in \eqref{eq:psi_final} is thus the trapping probability, and 
\begin{equation}\label{eq:findL}
    L=\frac{f(x')}{1-\psi(x')},
\end{equation}
for any $x'>1$, which completes the proof.
\end{proof}

\subsection{Trapping Probability Under Proportional Insurance Coverage} \label{AnalyticalApproachtoTrappingProbabilitiesviaInfinitesimalGenerators-Section4-Subsection2-Subsubsection2} 

\subsubsection{Analytic Solutions}\label{Trapping Probability Under Proportional Insurance Coverage - Analytic Solutions} 

In this section, we find explicit forms for a pair $u,w$ of linearly independent, real valued, solutions to $\mathcal{L}_\delta[h]=0$ with $\mathcal{L}_\delta$ given by \eqref{AnalyticalApproachtoTrappingProbabilitiesviaInfinitesimalGenerators-Section4-Equation23} 
that satisfy the conditions of Proposition \ref{prop:layer_method}.
The solutions are obtained in terms of hypergeometric functions. Indeed, recall that the hypergeometric differential equation with parameters $A, B$, and $C=1$ is given by (see, e.g., Equations (2.26) and (15.5.1) from \cite{Book:Seaborn1991} and \cite{Book:Abramowitz1964}, respectively):

\begin{align}
x(x-1) \frac{\rmd^2 h}{\rmd x^2} + ((A+B+1) x - 1) \frac{\rmd h}{\rmd x} + ABh = 0. 
\label{AnalyticalApproachtoTrappingProbabilitiesviaInfinitesimalGenerators-Section4-Equation44}
\end{align}

Its solutions have the form

\begin{align}
U(x) = x^{-A} ~ _2F_1\left(A,A,A-B+1,\frac{1}{x}\right), \quad 
    V(x) = x^{-B} ~ _2F_1\left(B,B,B-A+1,\frac{1}{x}\right) 
\label{eq:def_UV}
\end{align}

(see, e.g., Equations (15.5.7) and (15.5.8) from \cite{Book:Abramowitz1964}).

Clearly, the homogeneous equation $\mathcal{L}_\delta[h](x) = 0$ with $\mathcal{L}_\delta$ as in Equation \eqref{AnalyticalApproachtoTrappingProbabilitiesviaInfinitesimalGenerators-Section4-Equation23} is a hypergeometric differential equation with parameters $A, B,$ satisfying $A + B +1 = 2- \rho$, $A B  = \delta$. Hence, we define

\begin{align}
    A=\frac{1}{2}\left((1-\rho) + \sqrt{(1-\rho)^2 - 4 \delta}\right) \quad \text{ and } \quad
    B=\frac{1}{2}\left((1-\rho) - \sqrt{(1-\rho)^2 - 4 \delta}\right).
    \label{AnalyticalApproachtoTrappingProbabilitiesviaInfinitesimalGenerators-Section4-Equation46}
\end{align}

Our task now is to use the general solutions $U,V$ to construct real-valued solutions $u,w$ such that $u(1) = 1$, $w(1) = 0$ and $\tfrac{\rmd w}{\rmd s}(1^+)>0$, as per Proposition \ref{prop:layer_method}. The main tool for our derivation is Gauss' connection formula relating the behavior of $U$ at $x \approx 1$ to the values of hypergeometric functions near zero, in particular Equation \eqref{AppendixA:DefinitionandSelectedPropertiesofGaussHypergeometricFunction-Subsection1-Equation3} (see NIST Digital Library of Mathematical Functions, §15.8, Equation (15.8.1)),

\begin{align}
    U(x) &=  U_0 x^{-A} ~\!_2F_1\left(A,A,1-\rho,1-\frac{1}{x}\right) +
    U_\rho x^{-A} \left(1 - \frac{1}{x}\right)^\rho ~\!_2F_1\left(1-B,1-B,1+\rho,1-\frac{1}{x}\right)\nonumber \\
    &= U_0 u_0(x) + U_\rho u_\rho(x),
    \label{AnalyticalApproachtoTrappingProbabilitiesviaInfinitesimalGenerators-Section4-Equation47}
\end{align}
where we have used that $A+B = 1 -\rho$, and the coefficients are,
\begin{align}
    U_0 = U(1) = \frac{\Gamma(1+A-B)\Gamma(\rho)}{\Gamma(1-B)^2} \quad \text{ and } \quad
    U_\rho = \frac{\Gamma(1+A-B)\Gamma(-\rho)}{\Gamma(A)^2}.
    \label{AnalyticalApproachtoTrappingProbabilitiesviaInfinitesimalGenerators-Section4-Equation48}
\end{align}

The notation $u_0$ and $u_\rho$ used in Equation \eqref{AnalyticalApproachtoTrappingProbabilitiesviaInfinitesimalGenerators-Section4-Equation47} reflects the fact that these functions are precisely the two Frobenius branch solutions about the regular singular point $x=1$ of Equation \eqref{AnalyticalApproachtoTrappingProbabilitiesviaInfinitesimalGenerators-Section4-Equation44}, whose indicial exponents are $0$ and $\rho$.

The function $V$ can be similarly decomposed with $V_0 = V(1)$ and $V_\rho$ obtained by exchanging $A$ by $B$ in Equation \eqref{AnalyticalApproachtoTrappingProbabilitiesviaInfinitesimalGenerators-Section4-Equation48}. Further, note that we can apply Euler\rq s transformation formula \eqref{AppendixA:DefinitionandSelectedPropertiesofGaussHypergeometricFunction-Subsection1-Equation4} to $u_0$ and obtain

\begin{align}
    u_0(x) &= x^{-A} ~\!_2F_1\left(A,A,A+B,1-\frac{1}{x}\right) 
    = x^{-A}\left(\frac{1}{x}\right)^{B-A} ~\!_2F_1\left(B,B,A+B,1-\frac{1}{x}\right) \nonumber\\
    &= x^{-B}~\!_2F_1\left(B,B,1-\rho,1-\frac{1}{x}\right),
    \label{AnalyticalApproachtoTrappingProbabilitiesviaInfinitesimalGenerators-Section4-Equation49}
\end{align}

\begin{figure}
    \centering
    \includegraphics[scale=0.75]{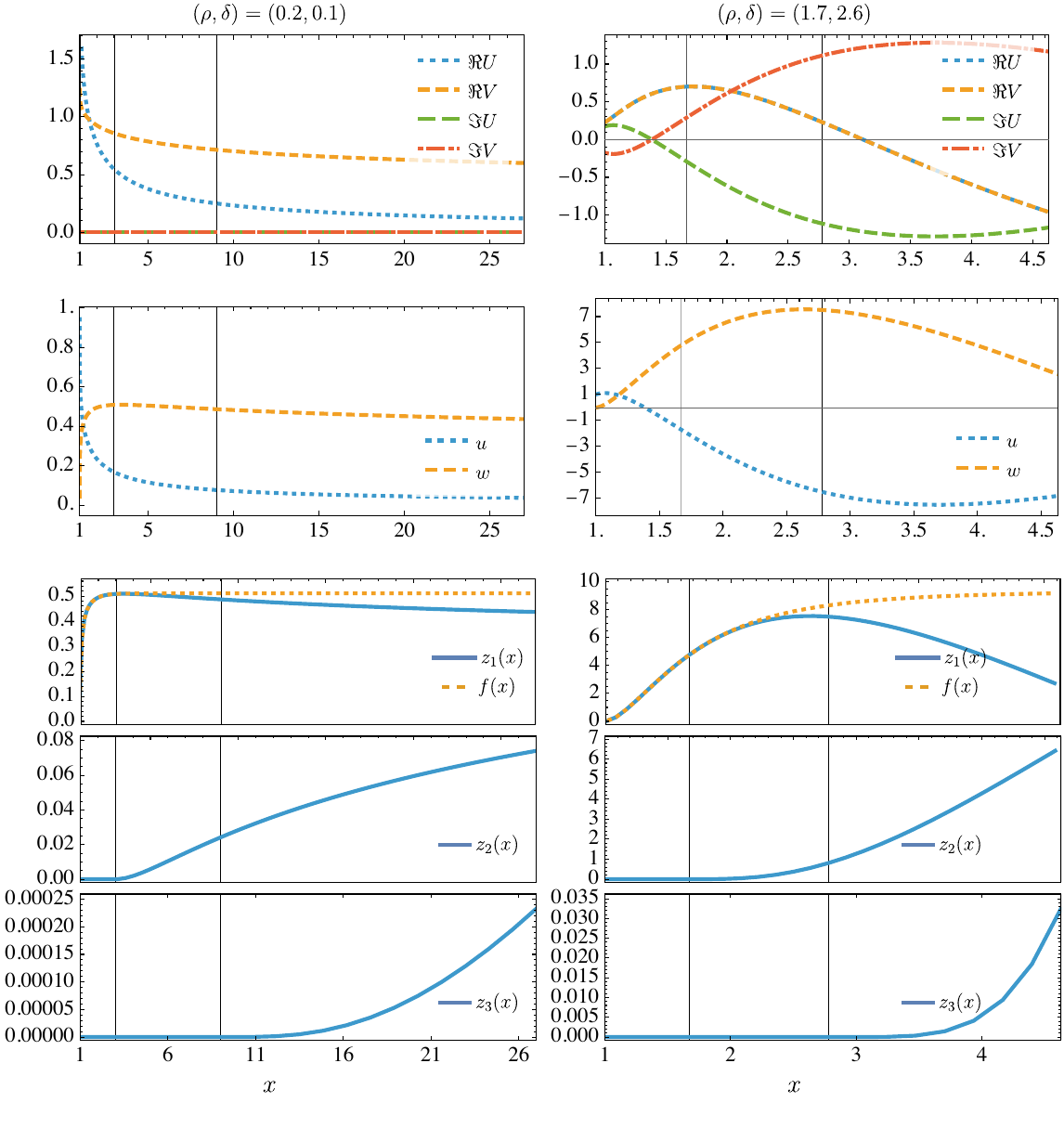}
    \caption{Illustration of the analytical method for the calculation of the trapping probability under proportional insurance as described in Section \ref{AnalyticalApproachtoTrappingProbabilitiesviaInfinitesimalGenerators-Section4-Subsection2}. The left column shows calculations for $(\rho,\delta) = (0.2,0.1)$ which corresponds to $(\kappa,\lambda) = (0.67,0.1)$ and $A,B$ being real-valued. The right column corresponds to the case labeled with the letter f in Figure \ref{fig:regions}. All plots have $x \in [1,x_3]$ as the independent variable, with the values of $x_1$ and $x_2$ marked with vertical lines. The functions $U,V$ in the first row are computed via Equation \eqref{eq:def_UV}. The second row shows the corresponding real solutions $u,w$ constructed in Propositions \ref{prop:uv_real} and \ref{prop:uv_complex}, respectively. The bottom three rows show the functions $z_1, z_2, z_3$ and $f$ constructed as per Proposition \ref{prop:layer_method}, and computed via numerical integration.
    }
    \label{fig:illustration}
\end{figure}

and similarly for $u_\rho$. This establishes two things. First, the function $V(x)$ can be decomposed as $V(x) = V_0 u_0(x) + V_\rho u_\rho(x)$. Secondly, in the case of imaginary $A$ and $B$, since $A = \bar{B}$, we have, by properties of the hypergeometric function and Equation \eqref{AnalyticalApproachtoTrappingProbabilitiesviaInfinitesimalGenerators-Section4-Equation49},

\begin{align}
    \overline{u_0(x)} = x^{-\bar{A}} ~\!_2F_1\left(\bar{A},\bar{A},1-\rho,1-\frac{1}{x}\right) = x^{-B}~\!_2F_1\left(B,B,1-\rho,1-\frac{1}{x}\right) = u_0(x).
    \label{AnalyticalApproachtoTrappingProbabilitiesviaInfinitesimalGenerators-Section4-Equation50}
\end{align}

Namely, $u_0$ is real-valued. And so is $u_\rho$ by an analogous calculation. Real-valued and complex cases of \(U\) and \(V\) are plotted in the first row of Figure \ref{fig:illustration}.
    
From the definition of $A,B$ in Equation \eqref{AnalyticalApproachtoTrappingProbabilitiesviaInfinitesimalGenerators-Section4-Equation46} and $U,V$ in Equation 
\eqref{eq:def_UV} we can see that the method of construction of the final set $\{u,w\}$ of linearly independent solutions to the homogeneous problem depends on the value of the discriminant $(1-\rho)^2 - 4 \delta$. Also, the qualitative behaviour of the solution depends on the sign of $\rho-1$. These regimes need to be compounded with the net profit condition in Equation \eqref{AnalyticalApproachtoTrappingProbabilitiesviaInfinitesimalGenerators-Section4-Equation51}.

Proposition \ref{prop:uv_real}
constructs the solutions $u$ and $w$ in the case where $U$ and $V$ are real.

\begin{proposition} \label{prop:uv_real}

Suppose $(1-\rho)^2>4 \delta$, corresponding to real solutions $U$ and $V$. Then the net profit condition holds only if $0<\rho<1$. In this regime, we define

\begin{align}
    u(x) = \frac{U(x)}{U(1)}, \quad v(x) = \frac{V(x)}{V(1)}, \quad \text{ and } \quad w(x) = v(x) - u(x).
    \label{AnalyticalApproachtoTrappingProbabilitiesviaInfinitesimalGenerators-Section4-Equation52}
\end{align}
    
Then $u,w$ are real valued, linearly independent solutions to the homogeneous problem \eqref{AnalyticalApproachtoTrappingProbabilitiesviaInfinitesimalGenerators-Section4-Equation31} satisfying $u(1)=1, w(1)=0$,
$\tfrac{\rmd w}{\rmd s}|_{1^+} >0,$ and $\tfrac{\rmd u}{\rmd s}|_{1^+}$ is finite.

\end{proposition}
\begin{proof}

We first prove that if $\rho >1$, the net profit condition forces the discriminant to be negative. Write the left hand side of Equation \eqref{AnalyticalApproachtoTrappingProbabilitiesviaInfinitesimalGenerators-Section4-Equation51} as $N_\rho(\delta)$ for $\delta > 0$ and note that $N_\rho$ is strictly decreasing on $(0,\infty)$ for any $\rho >0$. Evaluating at $\delta_0 := \frac{1}{4}(\rho-1)^2$ and using that $\ln(1+x) \leq x$ for all $x \geq 0$, we can bound

\begin{align}
    N_\rho(\delta_0) = \rho - \frac{(\rho - 1)^2}{2}\ln\left(1 + \frac{2}{\rho -1}\right) \geq \rho - (\rho -1) = 1.
    \label{ProofofProposition4.4-Equation1}
\end{align}

Since $N_\rho$ is decreasing, we have that $N_\rho(\delta) > 1$ for $\delta < \delta_0$. Namely, the net profit condition is violated whenever $\rho>1$ and $(1-\rho)^2>4 \delta$, as claimed.

Regarding $u,v$ and $w$, we note that clearly $u(1) = v(1) = 1$ and $w(1) = 0$. It suffices to prove $\frac{\rmd w}{\rmd s}(1^+)>0$. From continuation formula in Equation \eqref{AnalyticalApproachtoTrappingProbabilitiesviaInfinitesimalGenerators-Section4-Equation47},

\begin{align}
    w(x) &= \left(\frac{V_\rho}{V_0} - \frac{U_\rho}{U_0}\right) u_\rho(x) =
    \left(\frac{V_\rho}{V_0} - \frac{U_\rho}{U_0}\right) \left(1 - \frac{1}{x}\right)^\rho (1+O(x-1)) \quad \text{as } x\downarrow 1,
    \label{ProofofProposition4.4-Equation2}
\end{align}

which, since $0<\rho<1$, has infinite right derivative at $x=1$. However, differentiating by $\rmd s = \frac{\rmd x}{x(x-1)^{1-\rho}} $, we obtain:

\begin{align}
    \frac{\rmd w}{\rmd s}(x) = \frac{\rho}{x^\rho} \left(\frac{V_\rho}{V_0} - \frac{U_\rho}{U_0}\right) + O(x-1) , \quad \text{as } x \downarrow 1.
    \label{ProofofProposition4.4-Equation3}
\end{align}

Therefore, taking the limit and using Euler\rq s reflection formula yields, 

\begin{align}
    \frac{\rmd w}{\rmd s}(1^+) =& \rho \left(\frac{V_\rho}{V_0} - \frac{U_\rho}{U_0}\right)
    =\frac{\rho\Gamma(-\rho)}{\Gamma(\rho)}
    \left(\frac{\Gamma(1-A)^2}{\Gamma(B)^2} - \frac{\Gamma(1-B)^2}{\Gamma(A)^2}\right)\\
    =& \frac{-\Gamma(1-\rho)}{\Gamma(\rho)} \frac{(\Gamma(1-A)\Gamma(A))^2 -(\Gamma(1-B)\Gamma(B))^2}{(\Gamma(A)\Gamma(B))^2}\\
=&  \frac{\pi^3}{(\Gamma(A)\Gamma(B) \Gamma(\rho))^2}
\frac{\sin^2(\pi B)-\sin^2(\pi A)}{(\sin(\pi A)\sin(\pi B))^2 \sin(\pi \rho)}\\
=&  \frac{\pi^3}{(\Gamma(A)\Gamma(B) \Gamma(\rho))^2}
\frac{\sin(\pi (A+B))\sin(\pi (A-B))}{(\sin(\pi A)\sin(\pi B))^2 \sin(\pi \rho)}.
\label{ProofofProposition4.4-Equation4}
\end{align}

Since $\sin(\pi(A+B)) = \sin(\pi(1-\rho)) = \sin(\pi \rho)$ we obtain:

\begin{align}
    \frac{\rmd w}{\rmd s}(1^+) = \frac{\pi^3}{(\Gamma(A)\Gamma(B) \Gamma(\rho)\sin(\pi A)\sin(\pi B))^2}
\sin(\pi \sqrt{(1-\rho)^2-4 \delta}),
\label{ProofofProposition4.4-Equation5}
\end{align}

which is positive because $0<\rho<1$.

The finiteness of $\frac{\rmd u}{\rmd s}(1^+)$ follows similar arguments since, by \eqref{AnalyticalApproachtoTrappingProbabilitiesviaInfinitesimalGenerators-Section4-Equation47}, $u(x)= U_0 u_0(x) + U_\rho u_\rho(x) $, where $u_0(x)$ is clearly differentiable at 1.
\end{proof}

In the case with $(1-\rho)^2<4 \delta$, both $U$ and $V$ take complex values and $U(x) = \overline{V(x)}$. Then, the linearly independent solutions $\{u,w\}$ can be obtained by normalising the real and imaginary parts of $U$. This is detailed in Proposition \ref{prop:uv_complex}. The contrast between the cases with real and complex solution is illustrated in Figure \ref{fig:illustration}.

\begin{proposition} \label{prop:uv_complex}

Suppose $(1-\rho)^2 < 4 \delta$, let $U$ as in Equation 
\eqref{eq:def_UV}, and define 

\begin{align}
        v(x) = \begin{cases}
            \displaystyle{\frac{\Re U(x)}{\Re U(1)}}, & \text{if} \; \Re U(1)\Im U(1) >0 \\[2ex]
            \displaystyle{\frac{\Im U(x)}{\Im U(1)}}, & \text{if} \; \Re U(1)\Im U(1) <0 
        \end{cases}, \quad 
        u(x) = \begin{cases}
            \displaystyle{\frac{\Im U(x)}{\Im U(1)}}, & \text{if} \; \Re U(1)\Im U(1) >0 \\[2ex]
            \displaystyle{\frac{\Re U(x)}{\Re U(1)}}, & \text{if} \; \Re U(1)\Im U(1) <0 
        \end{cases},
        \label{AnalyticalApproachtoTrappingProbabilitiesviaInfinitesimalGenerators-Section4-Equation53}
\end{align}

and $w(x) = v(x) - u(x)$. Then $u,w$ are real valued, linearly independent solutions to the homogeneous problem \eqref{AnalyticalApproachtoTrappingProbabilitiesviaInfinitesimalGenerators-Section4-Equation31} satisfying $u(1)=1, w(1)=0,$ $\tfrac{\rmd w}{\rmd s}|_{1^+} >0,$ and $\tfrac{\rmd u}{\rmd s}|_{1^+}$ is finite.

\end{proposition}

\begin{proof}

Again, it suffices to prove $\frac{\rmd w}{\rmd s}(1^+)>0$. From the representation of $U(x)$ in Equation \eqref{AnalyticalApproachtoTrappingProbabilitiesviaInfinitesimalGenerators-Section4-Equation47} and recalling that $u_0$ and $u_\rho$ are real-valued, and that $U_0 = U(1)$, we can write

\begin{align}
    \frac{\rmd}{\rmd s} \left(\frac{\Re U(x)}{\Re U(1)} - \frac{\Im U(x)}{\Im U(1)} \right)
    &= \left( \frac{\Re U_\rho}{\Re U_0} - \frac{\Im U_\rho}{\Im U_0}\right) \frac{\rmd u_\rho}{\rmd s}(x).
    \label{ProofofProposition4.5-Equation1}
\end{align}

But $u_\rho(x) = (1-x)^\rho(1+O(x-1))$ as $x \downarrow 1$, then

\begin{align}
        \frac{\rmd u_\rho}{\rmd s}(x) = x(x-1)^{1-\rho}\left(\rho(x-1)^{\rho-1} + O((x-1)^\rho)\right) = \rho + O(x-1) \quad \text{as } x \downarrow 1.
        \label{ProofofProposition4.5-Equation2}
\end{align}

We thus have established that,

\begin{align}
        \frac{\rmd}{\rmd s} \left.\left(\frac{\Re U(x)}{\Re U(1)} - \frac{\Im U(x)}{\Im U(1)}\right) \right|_{x = 1^+} = \rho \left(\frac{\Re U_\rho}{\Re U_0} - \frac{\Im U_\rho}{\Im U_0} \right).
        \label{ProofofProposition4.5-Equation3}
\end{align}

Expanding $U_\rho$ and $U_0$ from Equation \eqref{AnalyticalApproachtoTrappingProbabilitiesviaInfinitesimalGenerators-Section4-Equation46} in polar form, we can write

\begin{align}
    \frac{\rho U_\rho}{U_0} = \rho \frac{\Gamma(-\rho)}{\Gamma(\rho)}
    \left| \frac{\Gamma(A+\rho)}{\Gamma(A)}\right|^2 e^{i \theta} 
    &= -\frac{\Gamma(1-\rho)}{\Gamma(\rho)}
        \left| \frac{\Gamma(A+\rho)}{\Gamma(A)}\right|^2 e^{i \theta}\\
    &= -\frac{\pi}{\Gamma(\rho)^2 \sin(\pi \rho)}
    \left| \frac{\Gamma(A+\rho)}{\Gamma(A)}\right|^2 e^{i \theta},
        \label{ProofofProposition4.5-Equation4}
\end{align}

where $\theta = 2(\arg \Gamma(A+\rho) - \arg \Gamma(A))$. It follows that the derivative in Equation \eqref{ProofofProposition4.5-Equation3} is

\begin{align}
        \frac{\Re (\rho U_\rho)}{\Re U_0} - \frac{\Im (\rho U_\rho)}{\Im U_0}  = 
        \left| \frac{\Gamma(A+\rho)}{\Gamma(A) \Gamma(\rho)}\right|^2  \frac{\pi |U_0|^2}{\Re U_0 \Im U_0} \frac{\sin(\theta)}{\sin(\pi \rho)}.
        \label{ProofofProposition4.5-Equation5}
\end{align}

Hence, in order to prove $\frac{\rmd w}{\rmd s}(1^+)>0$ for all cases, we just need to prove that $\frac{\sin(\theta)}{\sin(\pi \rho)} > 0$.

First, note that $A+\rho = 1- \bar{A}$, therefore
    
\begin{align}
    \theta &= 2(\arg \Gamma(1-\bar{A}) - \arg \Gamma(A)) = 2(\arg \overline{\Gamma(1-A)} - \arg \Gamma(A))\\
    &= -2 (\arg(\Gamma(1-A) \Gamma(A)) = -2 \arg\left(\frac{\pi}{\sin(\pi A)} \right)\\
    &= 2 \arg(\sin(\pi A)).
        \label{ProofofProposition4.5-Equation6}
\end{align}

Writing $\pi A = \alpha + i \beta$ with $\alpha = \frac{\pi}{2}(1-\rho)$, $\beta = \frac{\pi}{2}\sqrt{4 \delta- (1-\rho)^2} > 0$, we can express

\begin{align}
    \sin(\pi A) = \sin \alpha \cosh \beta + i \cos \alpha \sinh \beta, \quad
        \label{ProofofProposition4.5-Equation7}
\end{align}

and using properties of trigonometric functions, we derive

\begin{align}
        \sin(\theta) = \sin(2 \arg(\sin(\pi A)) \
        &= \frac{2 \sin \alpha \cos \alpha \cosh \beta \sinh \beta}{\sin^2 \alpha \cosh^2 \beta + \cos^2 \alpha \sinh^2 \beta}\\
        &=\frac{\sin(2 \alpha) \sinh(2 \beta)}{\cosh(2 \beta) - \cos(2\alpha)}.
        \label{ProofofProposition4.5-Equation8}
\end{align}

Noting that by definition of $\alpha$, $\sin(2 \alpha) = \sin(\pi \rho)$ and $\cos(2 \alpha) = - \cos(\pi \rho)$, then

\begin{align}
        \frac{\sin (\theta)}{\sin(\pi \rho)} = \frac{\sinh(2 \beta)}{\cosh(2 \beta) + \cos(\pi \rho)},
        \label{ProofofProposition4.5-Equation9}
\end{align}

which is always positive since $\beta>0$ implies $\sinh(2 \beta) >0 $ and $\cosh(2 \beta)>1$.

The same argument used in the proof of Proposition \ref{prop:uv_complex} to show that $\frac{\rmd u}{\rmd s}(1^+)$ is finite can be applied here since $u$ is, also in this case, given by \eqref{AnalyticalApproachtoTrappingProbabilitiesviaInfinitesimalGenerators-Section4-Equation47}.
\end{proof}

\setcitestyle{numbers} 
\bibliographystyle{chicago} 
\bibliography{arXiV_Submitted_240926}

\setcitestyle{authoryear}

\end{document}